\documentclass[a4paper, conference]{IEEEtran}

\usepackage{cite}
\usepackage{amsmath,amssymb,amsfonts}
\usepackage{algorithmic}
\usepackage{graphicx}
\usepackage{textcomp}
\usepackage{xcolor}
\usepackage{paralist}
\usepackage{eurosym}
\usepackage{flushend}
\usepackage{url}
\usepackage[utf8]{inputenc}
\usepackage[colorlinks=true]{hyperref}
\usepackage{soul}
\usepackage{caption}
\usepackage[list=true]{subcaption}
\usepackage{booktabs}
\usepackage{tabu}
\usepackage{wrapfig}
\usepackage{arydshln}
\usepackage[normalem]{ulem}
\usepackage{multirow}
\usepackage{siunitx}
\usepackage{adjustbox}
\usepackage{array}
\usepackage{diagbox}

\begin{document}

\title{GNSS Jammer Direction Finding in Dynamic Scenarios Using an Inertial-based Multi-Antenna System}

\author{\IEEEauthorblockN{Lucas Heublein\IEEEauthorrefmark{1}, Thorsten Nowak\IEEEauthorrefmark{2}, Tobias Feigl\IEEEauthorrefmark{1}, Jaspar Pahl\IEEEauthorrefmark{1}, \underline{Felix Ott}\IEEEauthorrefmark{1}}
  \IEEEauthorblockA{\IEEEauthorrefmark{1}Fraunhofer Institute for Integrated Circuits IIS, 90411 Nürnberg, Germany}
  \IEEEauthorblockA{\IEEEauthorrefmark{2}Diehl Defence GmbH \& Co. KG, 90552 Röthenbach an der Pegnitz, Germany}
  \IEEEauthorblockA{\{lucas.heublein, tobias.feigl, jaspar.pahl, felix.ott\}@iis.fraunhofer.de}
  \IEEEauthorblockA{thorsten.nowak@diehl-defence.com}
}

\IEEEoverridecommandlockouts
\IEEEpubid{\makebox[\columnwidth]{
979-8-3315-9568-5/25/\$31.00~\copyright2025
IEEE \hfill} \hspace{\columnsep}\makebox[\columnwidth]{ }}

\maketitle

\begin{abstract}
Jamming devices disrupt signals from the global navigation satellite system (GNSS) and pose a significant threat by compromising the reliability of accurate positioning. Consequently, the detection and localization of these interference signals are essential to achieve situational awareness, mitigating their impact, and implementing effective countermeasures. In this paper, we utilize a two-times-two patch antenna system (i.e., the software defined radio device Ettus USRP X440) to predict the angle, elevation, and distance to the jamming source based on in-phase and quadrature (IQ) samples. We propose to use an inertial measurement unit (IMU) attached to the antenna system to predict the relative movement of the antenna in dynamic scenarios. We present a synthetic aperture system that enables coherent spatial imaging using platform motion to synthesize larger virtual apertures, offering superior angular resolution without mechanically rotating antennas. While classical angle-of-arrival (AoA) methods exhibit reduced accuracy in multipath environments due to signal reflections and scattering, leading to localization errors, we utilize a methodology that fuses IQ and Fast Fourier Transform (FFT)-computed spectrograms with 22 AoA features and the predicted relative movement to enhance GNSS jammer direction finding.
\end{abstract}
\begin{IEEEkeywords}
  Global Navigation Satellite System, Jammer Localization, Angle of Arrival, Direction of Arrival, IQ Components, FFT, Machine Learning, Inertial Data\footnote{IEEE DGON Inertial Sensors and Applications (ISA), October 21-22 2025, Braunschweig, Germany}
\end{IEEEkeywords}
\IEEEpeerreviewmaketitle

\section{Introduction}
\label{label_introduction}

The localization accuracy of GNSS receivers is significantly compromised by interference signals emitted from jamming devices~\cite{ott_heublein_icl,heublein_raichur_ion,heublein_feigl_posnav,gaikwad_heublein}. This problem has intensified in recent years, primarily due to the widespread availability of inexpensive jammers. Consequently, it is imperative to either mitigate the impact of such interference or eliminate its source altogether. Effective countermeasure implementation necessitates the detection, classification~\cite{raichur_heublein}, and precise localization of the interference source.

Traditional jammer localization methodologies -- collectively referred to as direction-finding techniques -- include approaches such as Received Signal Strength (RSS), Angle of Arrival (AoA)~\cite{schmidt,zhu_chen_yang}, Direction of Arrival (DoA)~\cite{papageorgiou_sellathurai}, Time Difference of Arrival (TDoA), and Frequency Difference of Arrival (FDoA)~\cite{qiao_lu_lin}. In AoA and DoA-based methods, the incident direction of the interference signal at the receiver is estimated~\cite{yardibi_li_stoica}. However, these techniques are notably vulnerable to multipath effects, where reflected signals from structures such as buildings create multiple propagation paths. Such reflections can distort angle estimations, resulting in localization errors~\cite{heublein_feigl_crpa}. Furthermore, achieving high angular precision typically demands extensive antenna arrays and advanced hardware architectures, which substantially increase system cost and complexity.

Machine learning (ML) approaches have emerged as a promising solution to overcome the inherent shortcomings of conventional AoA and DoA techniques in GNSS interference localization. ML models are capable of learning complex spatial and statistical patterns from data, enabling them to effectively characterize and mitigate multipath propagation effects~\cite{heublein_feigl_crpa}. By compensating for signal distortions induced by reflections, these methods enhance the precision of angle estimation. Moreover, ML-based frameworks can identify obstructed propagation environments and adapt their estimations to maintain robustness under non-line-of-sight (NLoS) conditions~\cite{feintuch_tabrikian}. Recent research efforts increasingly emphasize hybrid architectures that integrate traditional AoA techniques with ML-based inference models~\cite{papageorgiou_sellathurai,zeng_gong_liu,nguyen_noubir} to leverage the complementary strengths of both domains. The objective of this work is to develop a framework for the localization of interference sources in challenging NLoS scenarios dominated by multipath effects.

In this study, we consider a scenario involving a mobile antenna receiver and a stationary interference source (jammer). The primary objective is to estimate the direction from the moving antenna to the jammer with high accuracy. Unlike conventional static-array approaches, the mobility of the antenna introduces an additional temporal dimension that can be exploited to enhance localization performance. By leveraging the antenna’s trajectory over time, it is possible to construct a \textit{synthetic aperture}, effectively increasing the spatial diversity of received signals without requiring a physically large antenna array~\cite{lamountain_closas}. This synthetic aperture enables improved resolution in DoA estimation and greater robustness against multipath interference~\cite{ahmed_sokolova}. The proposed framework aims to utilize this motion-induced spatial diversity to achieve more precise jammer localization under realistic operating conditions~\cite{borio_gioia}. To facilitate accurate prediction of the antenna’s relative pose, an inertial measurement unit (IMU) can be integrated into the antenna system, providing continuous estimates of its orientation and motion dynamics~\cite{yaqoob_mannesson}.

\textbf{Contributions.} The primary objective of this study is to localize GNSS jamming devices by estimating their range, azimuth, and elevation through the fusion of IQ signal data with frequency-domain representations derived from FFT-based spectrograms. To achieve this, we propose a hybrid framework that integrates vision-encoder ML models with time-series architectures, augmented by classical features extracted from 22 conventional AoA techniques. The proposed fusion strategy incorporates a sequence of five temporal steps as model input to exploit temporal dependencies and enhance direction-finding accuracy. For rigorous evaluation, a large-scale benchmark dataset was collected at Fraunhofer IIS in Nürnberg under industrial conditions, comprising both stationary jamming devices and a software-defined radio (SDR) receiver mounted on a 3D positioning system. This setup enables dynamic trajectory measurements with ground-truth position data, explicitly capturing the impact of varying multipath propagation effects.

\textbf{Outlook.} The remainder of this paper is structured as follows. Section 2 presents literature on GNSS interference direction finding with synthetic apertures. Section 3 introduces the proposed fusion methodology. The recording setup and dataset employed in this study are described in Section 4, while Section 5 presents a summary of the evaluation results. Finally, Section 6 concludes.
\section{Related Work}
\label{label_related_work}

\subsection{Jammer Direction Finding}

Qiao et al.~\cite{qiao_lu_lin} provided an extensive review of methodologies for estimating amplitude, phase, and spatial spectra, focusing on their applications in interference source direction finding. Within this domain, the Multiple Signal Classification (MUSIC) algorithm~\cite{schmidt} has emerged as a leading technique for spatial spectrum estimation. Building on this foundation, Papageorgiou et al.~\cite{papageorgiou_sellathurai} introduced a convolutional neural network (CNN)-driven model that infers the DoA from an estimated sample covariance matrix. In a related effort, Nguyen et al.~\cite{nguyen_noubir} designed a general-purpose anti-jamming framework that employs CNNs to identify interference, quantify the number of active emitters, and estimate their phase characteristics. Yardibi et al.~\cite{yardibi_li_stoica} proposed an iterative adaptive approach (IAA) based on nonparametric least-squares optimization for simultaneous amplitude and phase estimation. Feintuch et al.~\cite{feintuch_tabrikian} further extended CNN-based DoA estimation to multisource environments, addressing cases where the number of emitters is not known in advance and the interference exhibits non-Gaussian behavior. In parallel, Zeng et al.~\cite{zeng_gong_liu} developed the Multi-Channel Attentive Feature Fusion (McAFF) architecture for radio frequency (RF) fingerprinting, which integrates neural features from diverse signal representations -- such as IQ samples, carrier frequency offsets, FFT coefficients, and short-time Fourier transform (STFT) spectra -- to enhance identification performance.

\subsection{Spatio-temporal Synthetic Aperture}

A comprehensive overview of synthetic aperture radar (SAR) system design, signal models, and the principal image-formation and processing algorithms used to create high-resolution synthetic-aperture images is given in~\cite{curlander_mcdonough}. Synthetic aperture gives spatial diversity (aperture baseline, multi-look/multistatic views) that separates angular signatures of jammers vs. targets. The temporal evolution of returns across the aperture (slow-time) encodes kinematic differences (e.g., fixed jammer vs. moving target) and coherence properties that decomposition/filtering methods can exploit. Combining spatial and temporal structure yields stronger separation between jammer and scene, which reduces false targets. Piecewise Sub-Aperture (STAP) \cite{shen_tang_nie} divides the synthetic aperture into short segments and applies space-time adaptive processing to exploit local stationarity for improved jammer direction and range estimation. Low-rank methods \cite{liu_xu_ding} model SAR data as the sum of structured (signal) and sparse (jammer) components, using their spatio-temporal differences to isolate and localize interference sources. Wang et al.~\cite{wang_wu_pei} proposed a collaborative multi-static fusion that combines observations from multiple synthetic apertures or receivers across time to triangulate jammer positions and suppress deceptive signals. Variable space-frequency filtering \cite{hendy_alhourani} applies adaptive filtering in the joint space–time (or space–frequency) domain to separate jammers by exploiting their distinct spatio-temporal signatures. Three main methods use IMU data to predict relative pose/egomotion and combine that pose with ML or data-driven methods to improve RF source localization or direction finding. Neural inertial localization (NILoc), proposed by Herath et al.~\cite{herath_caruso}, is a transformer-based deep model that turns raw IMU streams into velocity/pose estimates (relative trajectory) which can be fed to downstream ML localization to greatly reduce pose uncertainty when mapping RF measurements to world coordinates. Jiang et al.~\cite{jiang_caruso} presented a fusion approach that uses IMU-predicted relative pose together with radio ranging (UWB/Wi-Fi) and a recurrent/optimization backend to produce robust position estimates in NLOS/multipath conditions that are directly applicable to improving jammer/source geolocation. Santra et al.~\cite{santra_wang_shaker} proposed an IMU-derived pose alongside RF fingerprints (i.e., RSSI, ToA, and angle features) to disambiguate multipath and thereby improve direction finding and geolocation accuracy for emitters. However, there are relatively few paper that combine IMU specifically for jammer localization.
\section{Methodology}
\label{label_method}

\begin{figure*}[!t]
    \centering
    \includegraphics[width=1.0\linewidth]{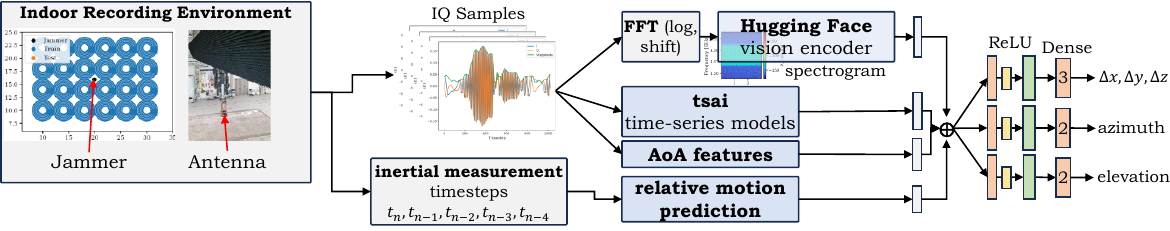}
    \caption{Overview of the proposed pipeline. Raw measurements are acquired using a dynamic Ettus USRP and subsequently converted into IQ samples. Vision encoders are trained on spectrograms derived via Fast Fourier Transform (FFT) and are integrated with time-series models and statistical feature representations. Furthermore, five measurement sets are processed to estimate the relative motion using data from an inertial sensor.}
    \label{figure_methodology}
    \vspace{-0.2cm}
\end{figure*}

\begin{figure}[!t]
    \centering
    \includegraphics[width=1.0\linewidth]{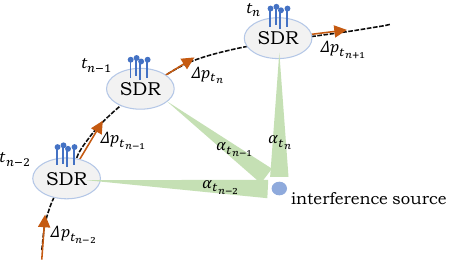}
    \caption{The figure illustrates successive positions of an SDR system with four antennas moving over time $t_{n-2}, t_{n-1}, t_n$ while measuring angle-of-arrival ($\alpha$) signals from an interference source.}
    \label{figure_setup_SDR}
    \vspace{-0.2cm}
\end{figure}

In this section, we present the experimental setup and methodology, as illustrated in Figure~\ref{figure_setup_SDR}. The system consists of an SDR platform equipped with an IMU, which moves through the environment and can be mounted on a mobile platform, such as a vehicle, in real-world applications. The objective is to estimate the direction of an interference source. At each discrete timestep $t_n$, the SDR platform records IQ samples while its relative position and orientation $\Delta p_{(t_n)}$ are obtained from the IMU.

As shown in Figure~\ref{figure_setup_SDR}, the SDR sequentially occupies positions corresponding to different timesteps $(t_{n-2}, t_{n-1}, t_n )$, forming a synthetic aperture as it moves. At each position, the SDR measures the angle of arrival ($\alpha_{t_n}$) of signals originating from the interference source. By combining the IQ data collected across multiple positions with the relative pose information from the IMU, the effective aperture is extended synthetically, thereby enhancing the direction-finding accuracy and robustness of the interference localization process. The objective is to integrate the IQ samples with the predicted relative poses to obtain direction finding results. Accordingly, we propose a fusion method architecture to achieve this goal.

An outline of the proposed methodology is presented in Figure~\ref{figure_methodology}. The process begins with the acquisition of a dataset comprising GNSS raw measurements collected using a mobile SDR operating within an indoor industrial environment equipped with a stationary jamming device. A 3D positioning system is employed to obtain precise spatial labels (see Section 4). Each dataset entry contains the raw measurements along with corresponding $x$-, $y$-, and $z$-coordinate labels relative to the position of the jamming source.

The acquired measurements are subsequently transformed into raw IQ samples for each antenna, which are processed through four independent analytical pipelines:
\begin{enumerate}
    \item Spectrogram-based vision encoding: The IQ samples are converted into spectrograms of dimensions $4 \times 512 \times 479$ using an FFT. These spectrograms are utilized to train a vision encoder sourced from the Hugging Face model repository~\cite{hugging_face}. The highest-performing model is then selected for feature fusion, followed by processing through a linear layer.
    
    \item Time-series modeling: The IQ samples are directly analyzed using 18 distinct time-series models from the \textit{tsai} library~\cite{tsai}. The optimal model, determined empirically, generates a feature representation of size 128, which is subsequently incorporated into the fusion process.
    
    \item AoA feature extraction: For each antenna, 22 conventional AoA estimation features are computed in accordance with the framework proposed by Wu et al.~\cite{wu_zhou_shen}. These features are further processed via a one-dimensional convolutional layer (kernel size = 1) and a linear layer, producing a final feature vector of dimension 32.
    
    \item An IMU mounted on the SDR platform is utilized to estimate the relative pose at each timestep. The resulting motion predictions serve as auxiliary inputs, which are concatenated with the outputs of the preceding three layers. Incorporating IMU-based relative pose information introduces complementary motion and orientation cues, enabling the model to differentiate between signal variations arising from platform dynamics and those caused by environmental interference, thereby improving the stability and physical consistency of the GNSS jammer localization process.
\end{enumerate}

The feature representations derived from all four processing paths are concatenated into a unified vector of size 288. For each regression task, this representation is passed through a dense layer comprising 512 units with ReLU activation, followed by a final dense layer (output size 2 or 3) that predicts the relative displacement $[\Delta x, \Delta y, \Delta z]$ between the antenna and the jamming device, as well as the azimuth ($\alpha$) and elevation ($\beta$) angles. A tanh activation function is applied prior to the computation of azimuth and elevation values. To prevent overfitting, dropout regularization with rates of 50\% is applied both before feature concatenation and after the final dense layer.

As a state-of-the-art baseline, our approach adopts the McAFF architecture \cite{zeng_gong_liu} for GNSS jammer localization. In this framework, each input IQ sample is processed through four parallel computational pathways, each designed to extract complementary signal characteristics: (1) Raw measurement pathway: The unprocessed IQ data are analyzed using a series of two-dimensional convolutional layers to capture spatial and temporal correlations. (2) Carrier frequency offset (CFO) pathway: The accumulated CFO is utilized to model phase variations between received signals, providing phase-domain information. (3) Frequency-domain pathway: The FFT coefficients are computed to obtain a compact frequency-domain representation of the signal spectrum. (4) Time-frequency pathway: The short-time Fourier Transform (STFT) coefficients, representing a sequence of discrete Fourier Transforms, are extracted and processed analogously to capture time-varying spectral behavior. The outputs from these four paths are fused via a shared attention module, enabling the network to emphasize the most informative features across modalities. The resulting representations are then concatenated and passed through a ResNeXt~\cite{xie_girshick} block, which performs high-level feature extraction. Finally, the network outputs predictions of distance, azimuth, and elevation, trained using the same loss function employed in our proposed method.
\section{Setup \& Dataset Recording}
\label{label_experiments}

\begin{figure}[!t]
    \centering
	\begin{minipage}[t]{0.492\linewidth}
        \centering
        \includegraphics[width=1.0\linewidth]{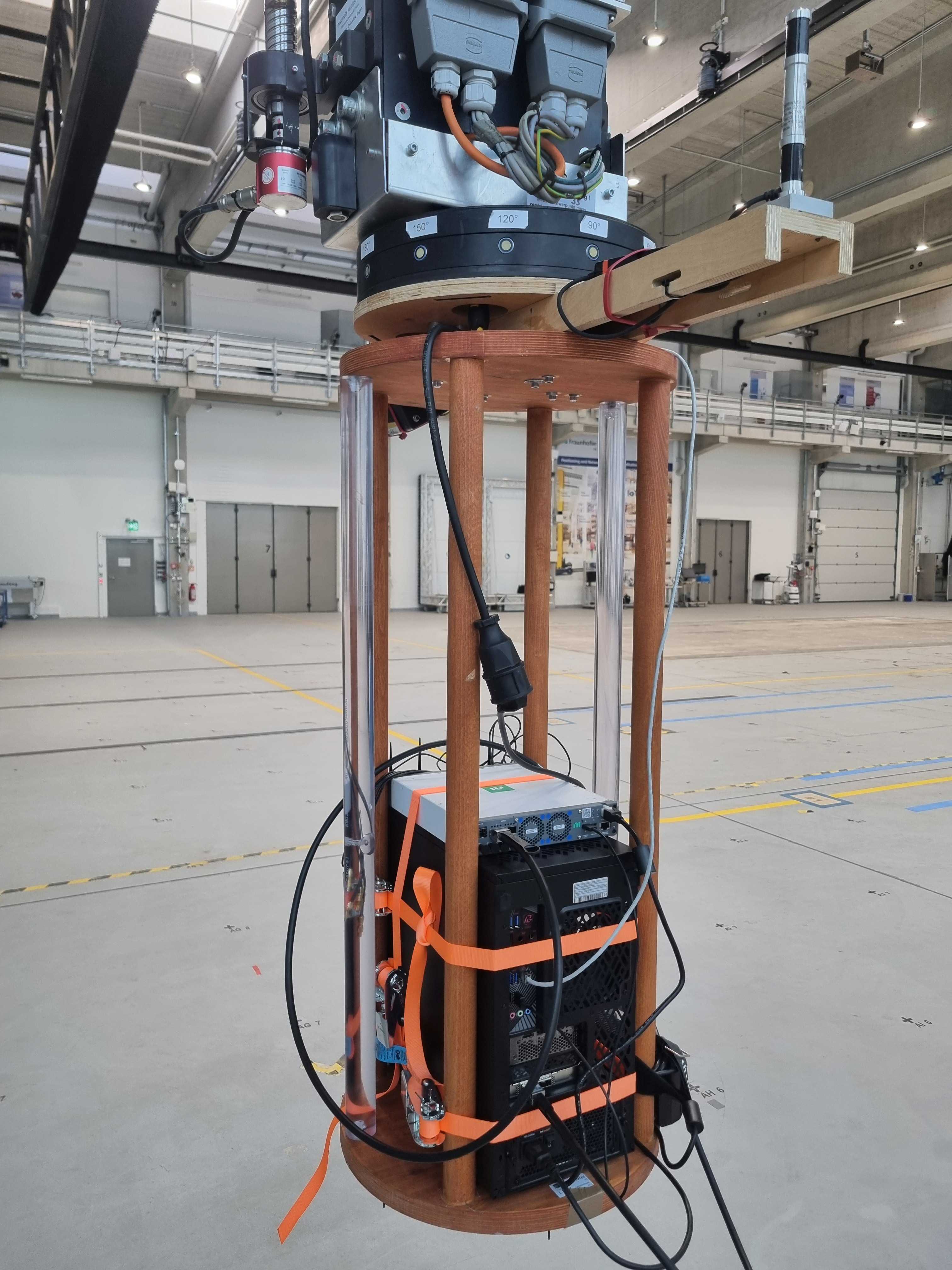}
        \caption{The recording setup includes an Ettus USRP X440 and a PC mounted at the 3D positioning system.}
        \label{figure_setup1}
    \end{minipage}
    \hfill
	\begin{minipage}[t]{0.492\linewidth}
        \centering
        \includegraphics[width=1.0\linewidth]{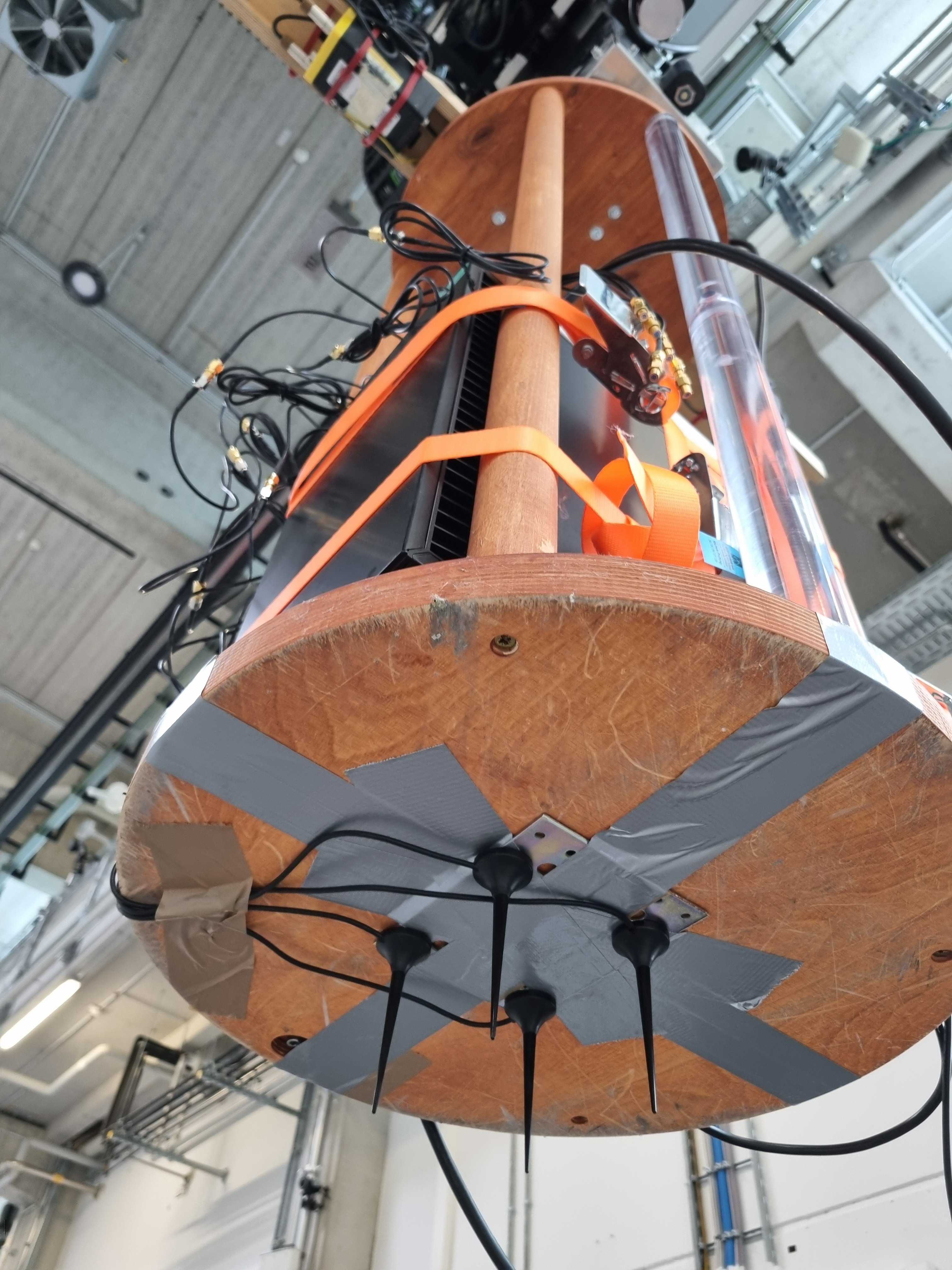}
        \caption{Placement of four sweep antennas mounted at the bottom of the positioning system with $9\,cm$ distance.}
        \label{figure_setup2}
    \end{minipage}
\end{figure}

To enable a comprehensive evaluation of the proposed models, a large-scale measurement campaign was conducted at the Fraunhofer IIS L.I.N.K.~test and application center. This facility provides a controlled industrial environment that allows the recording of arbitrary antenna trajectories with diverse motion patterns. The experimental setup employs a software-defined radio (SDR) platform mounted on a dynamic positioning system (see Section 4.1) to capture received signal data, while multiple stationary jamming devices serve as interference sources (see Section 4.2). The measurement environment, including its structural and electromagnetic characteristics, is detailed in Section 4.3. A summary of the resulting dataset, encompassing trajectory configurations, signal conditions, and annotation methodology, is provided in Section 4.4.

\subsection{Ettus USRP X440 Setup}

As the recording platform, we employ the Ettus USRP X440 SDR, which provides eight transmit (TX) and eight receive (RX) channels based on a direct-sampling transceiver architecture. The device integrates a Xilinx Zynq UltraScale+ RFSoC with a programmable FPGA, supporting the open-source UHD and RFNoC toolchains for flexible integration of external up- and down-conversion stages, filtering modules, and amplification front ends.

The X440 SDR supports an instantaneous bandwidth of up to 1.6\,GHz within an operational frequency range of 30\,MHz to 4\,GHz. Data streaming is realized via a dual 100\,Gb Ethernet (QSFP28) interface connected through a QSFP28-to-4$\times$SFP28 breakout cable to a high-performance host computer. The complete setup is illustrated in Figure~\ref{figure_setup1}. Four ANT-5GW-MMG2-SMA cellular antennas are mounted at the lower section of the positioning system, forming a quadratic array with a side length of $9\,cm$ (see Figure~\ref{figure_setup2}). The recording bandwidth is restricted to 40.96\,MHz with a center frequency of 1.57542\,GHz, and data are acquired in $3\,ms$ snapshots with an inter-frame interval of $200\,ms$, resulting in a total of five snapshots per second.

\begin{figure}[!t]
    \centering
	\begin{minipage}[t]{0.492\linewidth}
        \centering
        \includegraphics[width=1.0\linewidth]{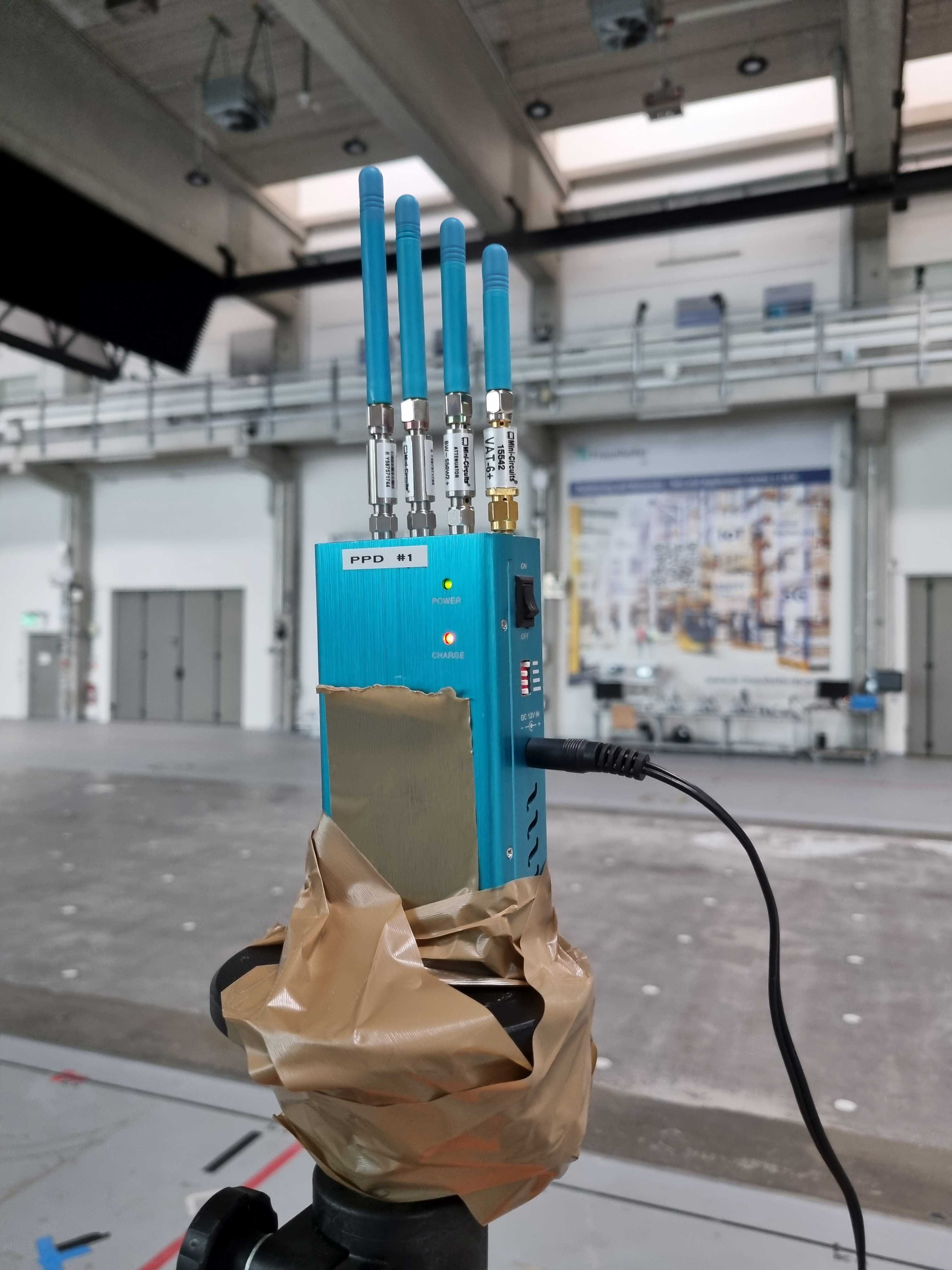}
        \caption{Jamming device 1 (blue).}
        \label{figure_jammer1}
    \end{minipage}
    \hfill
	\begin{minipage}[t]{0.492\linewidth}
        \centering
        \includegraphics[width=1.0\linewidth]{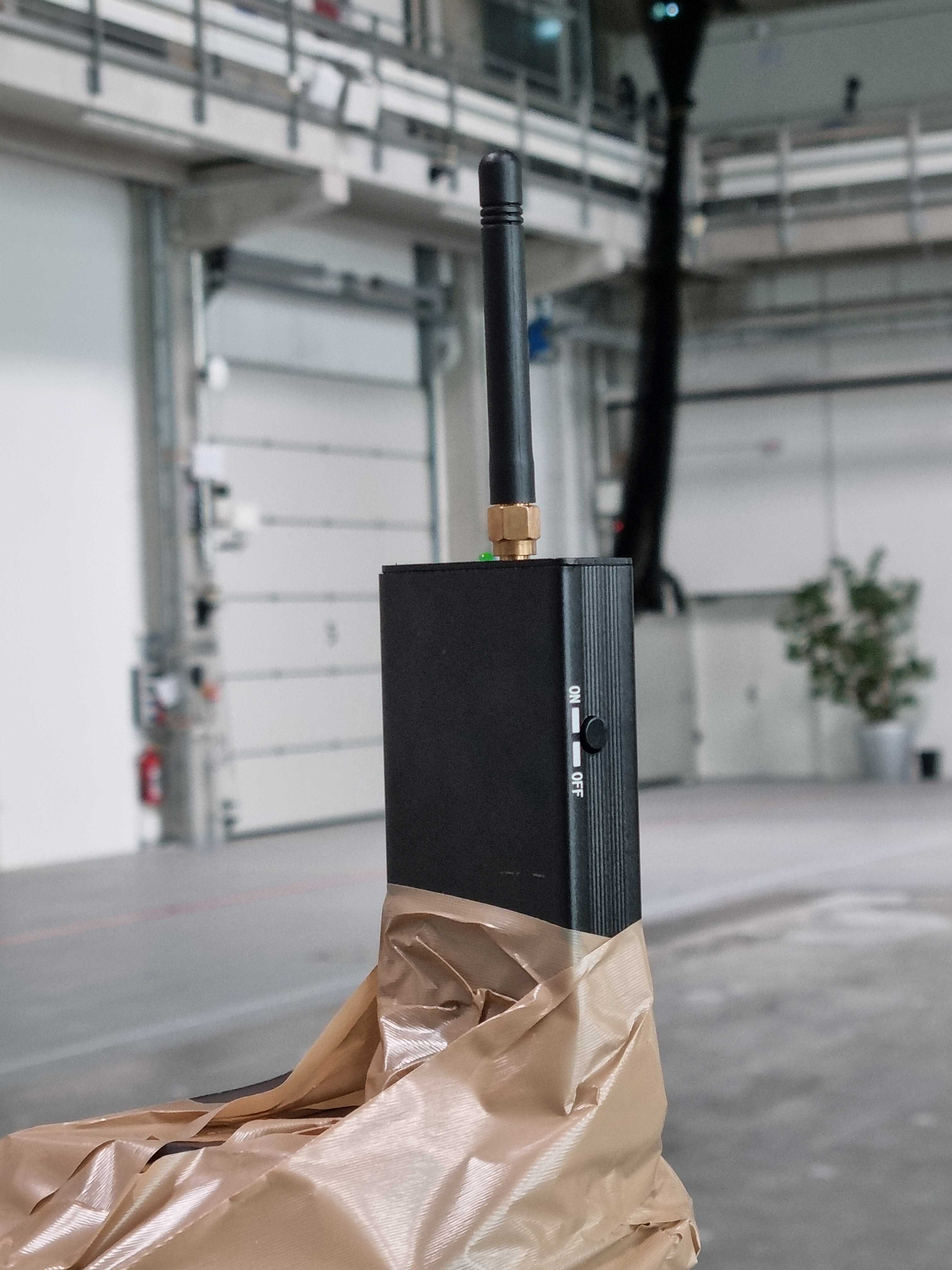}
        \caption{Jamming device 2 (black).}
        \label{figure_jammer2}
    \end{minipage}
\end{figure}

\begin{figure*}[!t]
    \centering
	\begin{minipage}[t]{0.492\linewidth}
        \centering
        \includegraphics[width=1.0\linewidth]{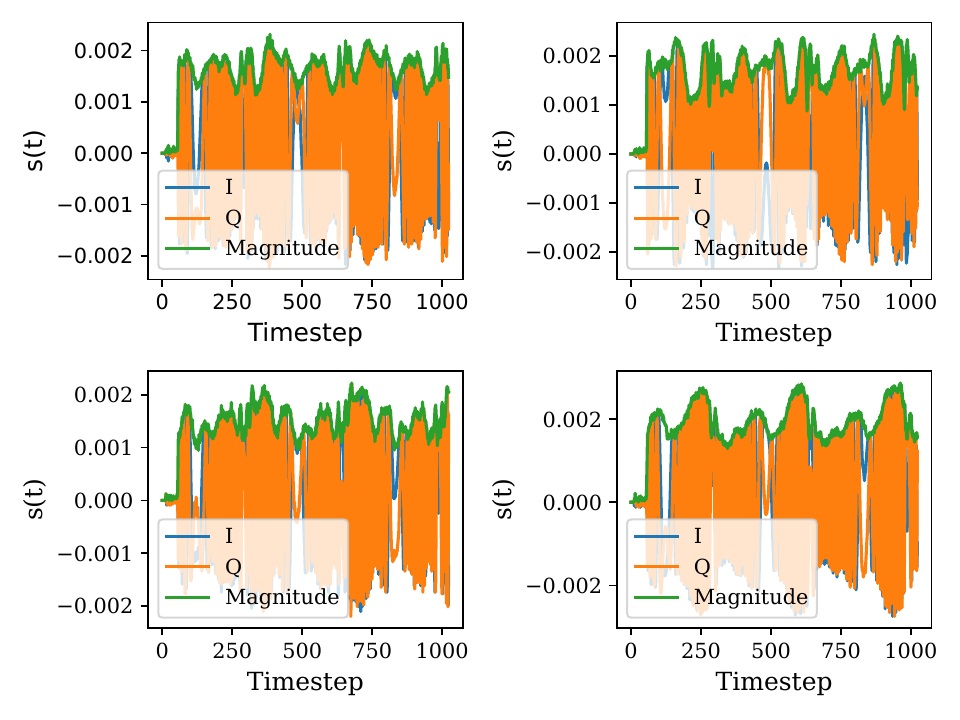}
        \caption{Plots of the imaginary, real, and magnitude for the first jammer (blue) for all four antennas.}
        \label{figure_iq1}
    \end{minipage}
    \hfill
	\begin{minipage}[t]{0.492\linewidth}
        \centering
        \includegraphics[width=1.0\linewidth]{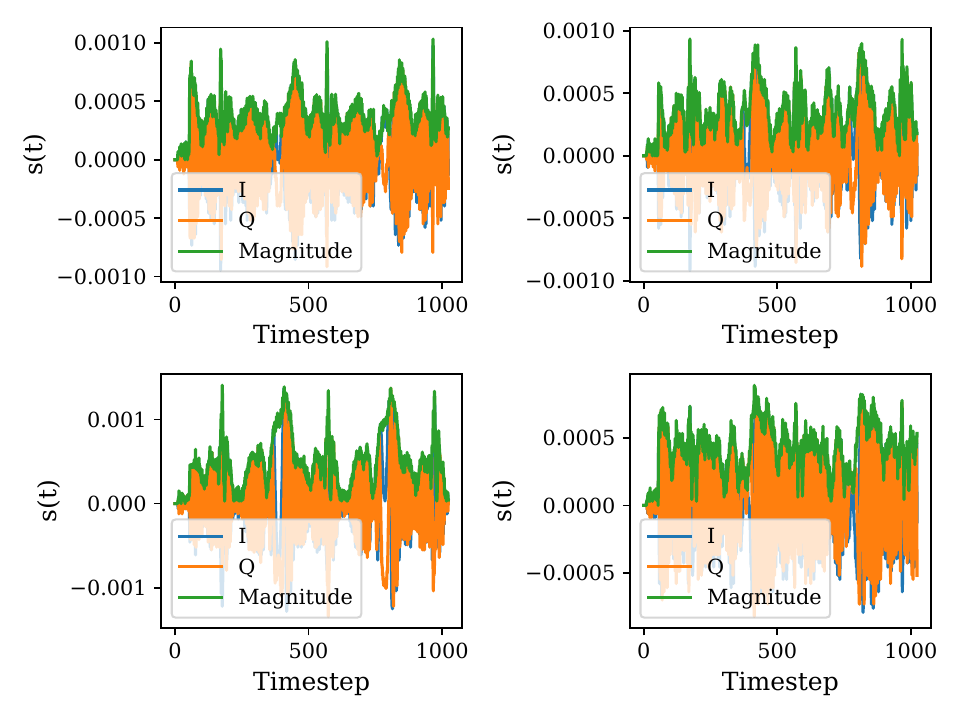}
        \caption{Plots of the imaginary, real, and magnitude for the second jammer (black) for all four antennas.}
        \label{figure_iq2}
    \end{minipage}
\end{figure*}

\subsection{Jamming Devices}

Two distinct privacy-preserving devices (PPDs) are employed as interference sources (see Figure~\ref{figure_jammer1} and Figure~\ref{figure_jammer2}). Both devices generate chirp-type interference signals with bandwidths of 20\,MHz (Figure~\ref{figure_jammer1}) and 25\,MHz (Figure~\ref{figure_jammer2}), respectively, and a nominal output power of 10\,dBm. The center frequency of the jamming signals is set to 1.57542\,GHz.

\subsection{L.I.N.K.~Measurement Environment}

\begin{figure}[!t]
    \centering
	\begin{minipage}[t]{0.492\linewidth}
        \centering
        \includegraphics[width=1.0\linewidth]{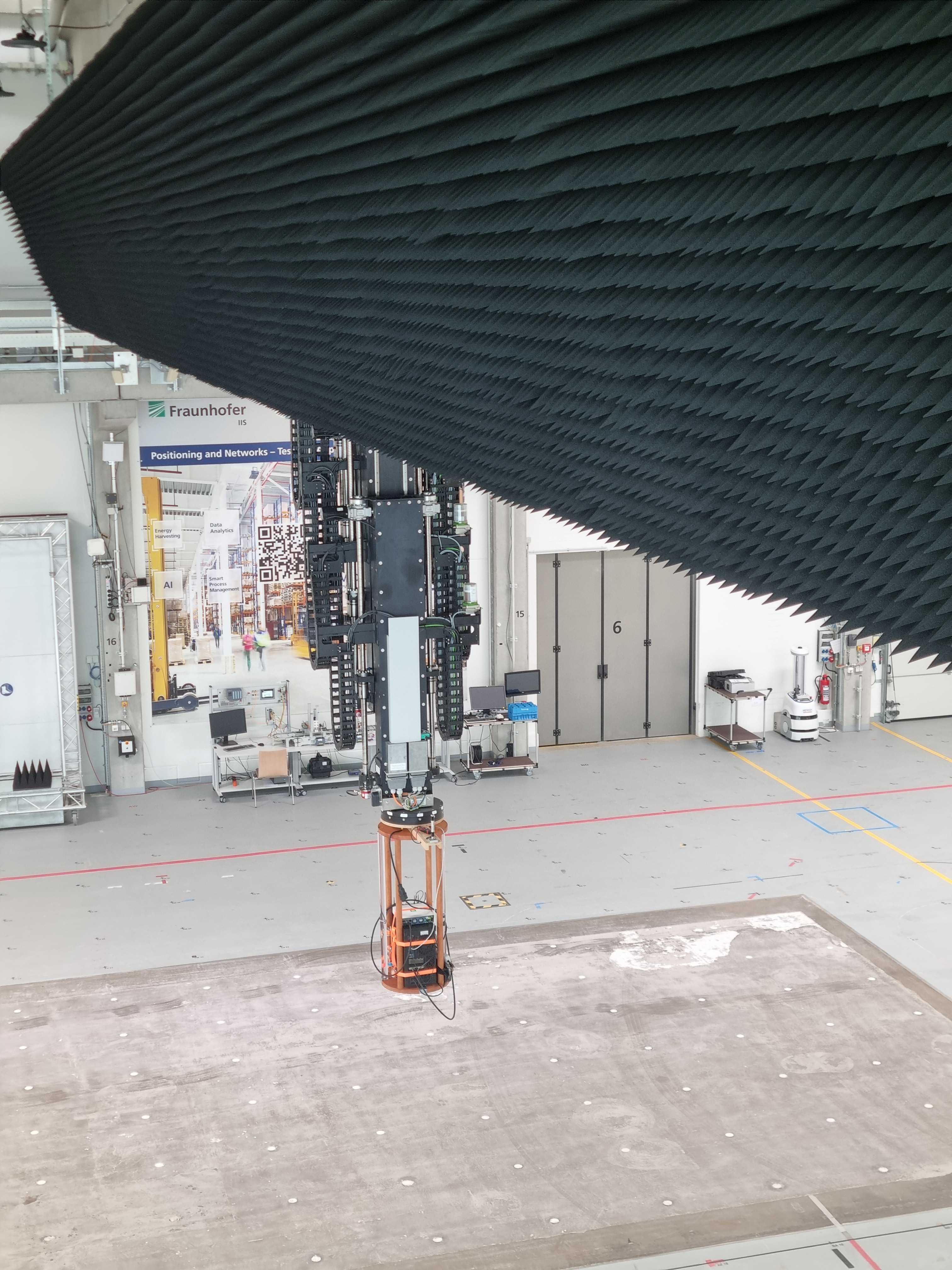}
        \caption{3D positioning system with mounted setup in the test and application center.}
        \label{figure_env1}
    \end{minipage}
    \hfill
	\begin{minipage}[t]{0.492\linewidth}
        \centering
        \includegraphics[width=1.0\linewidth]{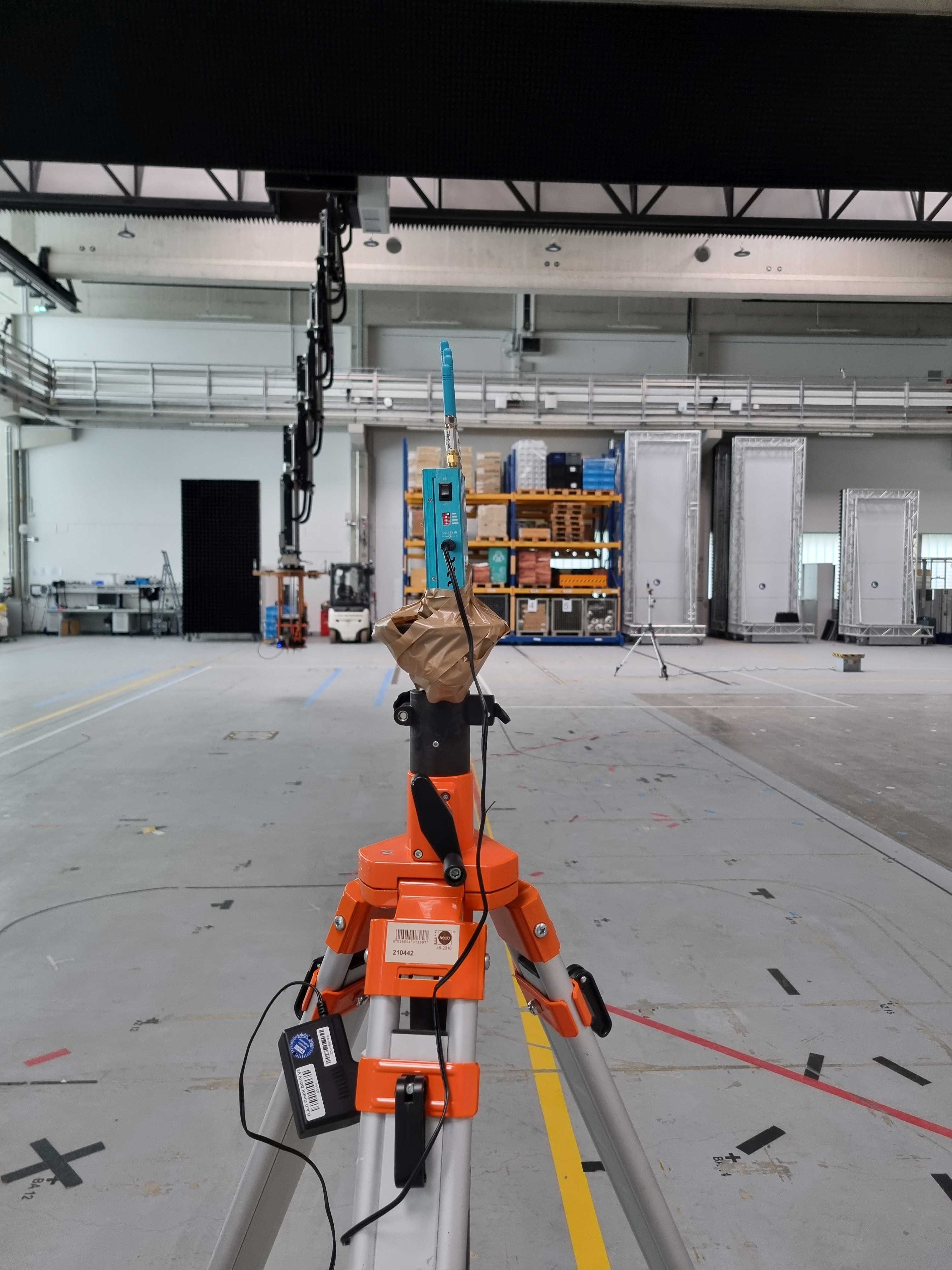}
        \caption{Two jamming devices stationary placed in the test and applications center.}
        \label{figure_env2}
    \end{minipage}
\end{figure}

Measurements are conducted in the L.I.N.K.~test and application center, which replicates a large-scale industrial environment characterized by complex multipath propagation (see Figure~\ref{figure_env1}). The 3D positioning system allows the receiver platform to traverse arbitrary locations within the environment, with a maximum velocity of $0.3\frac{m}{s}$. Stationary jamming devices are positioned near the center of the test area (see Figure~\ref{figure_env2}).

\begin{figure}[!t]
    \centering
	\begin{minipage}[t]{0.58\linewidth}
        \centering
        \includegraphics[width=1.0\linewidth]{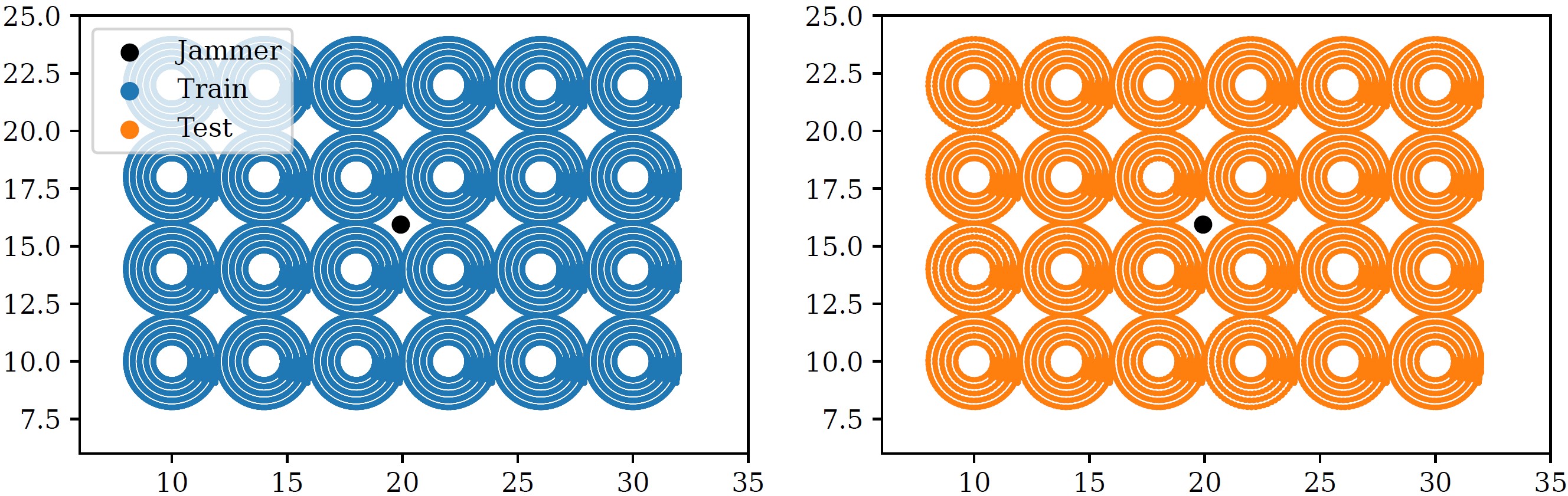}
        \caption{Train (left) and test (right) trajectory pattern for the first jamming device (black).}
        \label{figure_trajectory1}
    \end{minipage}
    \hfill
	\begin{minipage}[t]{0.38\linewidth}
        \centering
        \includegraphics[width=1.0\linewidth]{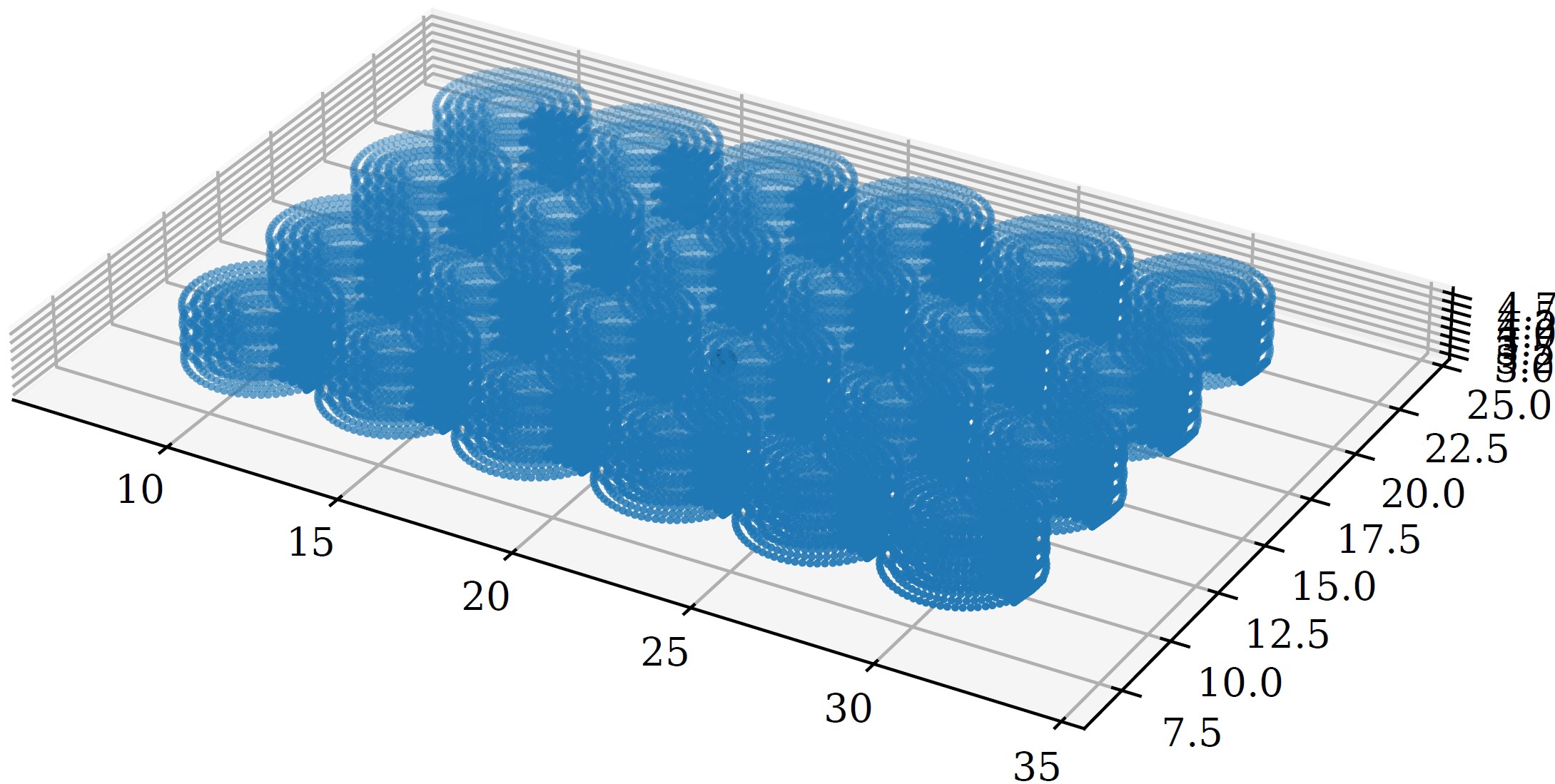}
        \caption{Trajectory pattern with different heights for the first jammer (black).}
        \label{figure_trajectory2}
    \end{minipage}
\end{figure}

\begin{figure}[!t]
    \centering
	\begin{minipage}[t]{0.58\linewidth}
        \centering
        \includegraphics[width=1.0\linewidth]{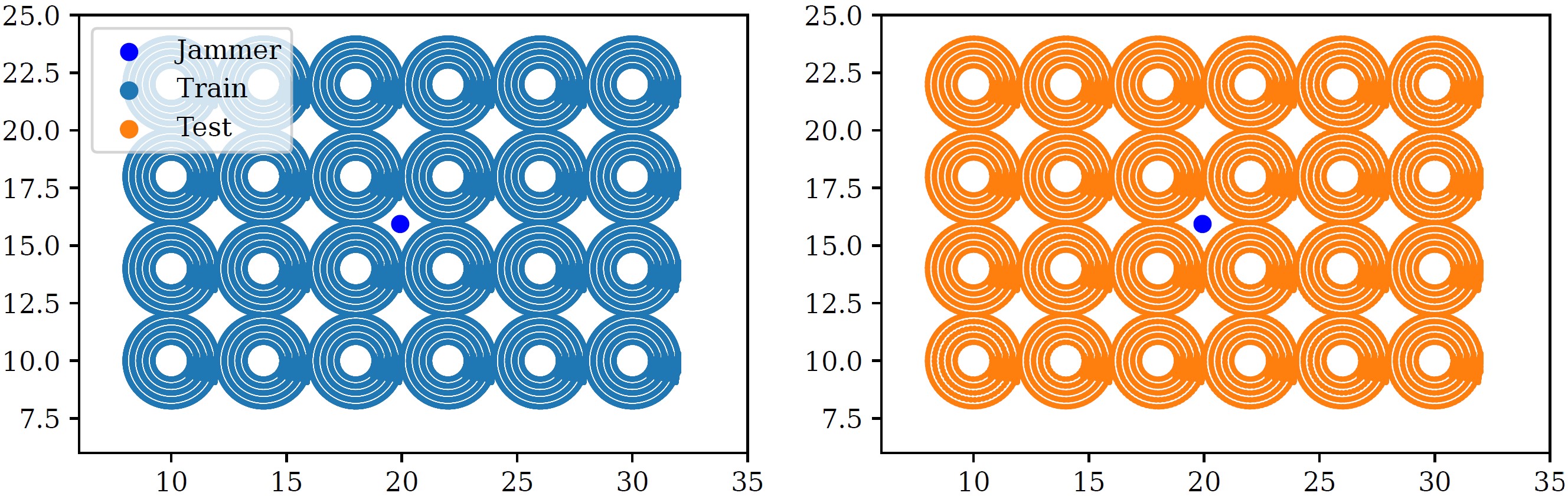}
        \caption{Train (left) and test (right) trajectory pattern for the second jamming device (blue).}
        \label{figure_trajectory3}
    \end{minipage}
    \hfill
	\begin{minipage}[t]{0.38\linewidth}
        \centering
        \includegraphics[width=1.0\linewidth]{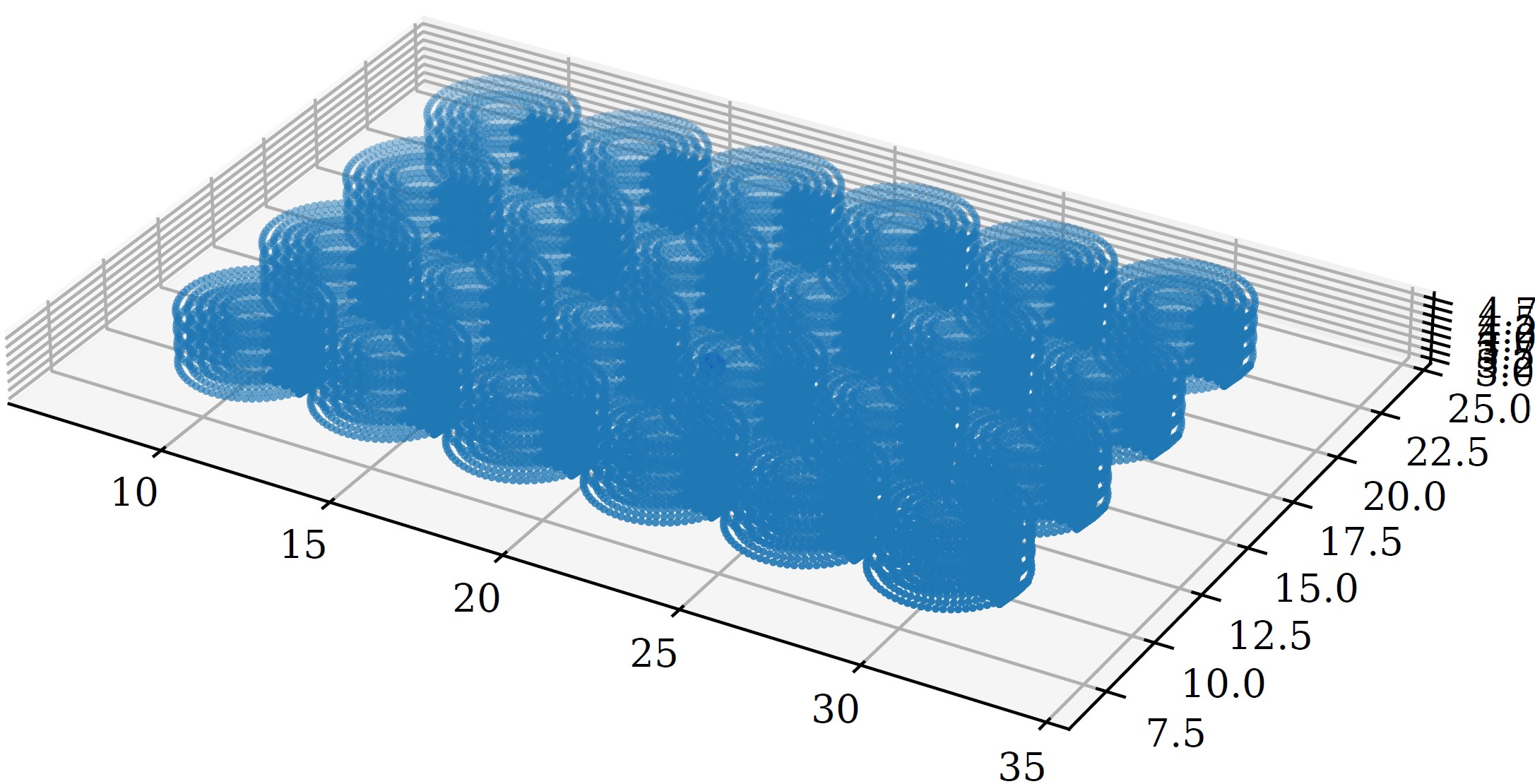}
        \caption{Trajectory pattern with different heights for the second jammer (blue).}
        \label{figure_trajectory4}
    \end{minipage}
\end{figure}

\begin{figure*}[!t]
    \centering
	\begin{minipage}[t]{0.492\linewidth}
        \centering
        \includegraphics[width=1.0\linewidth]{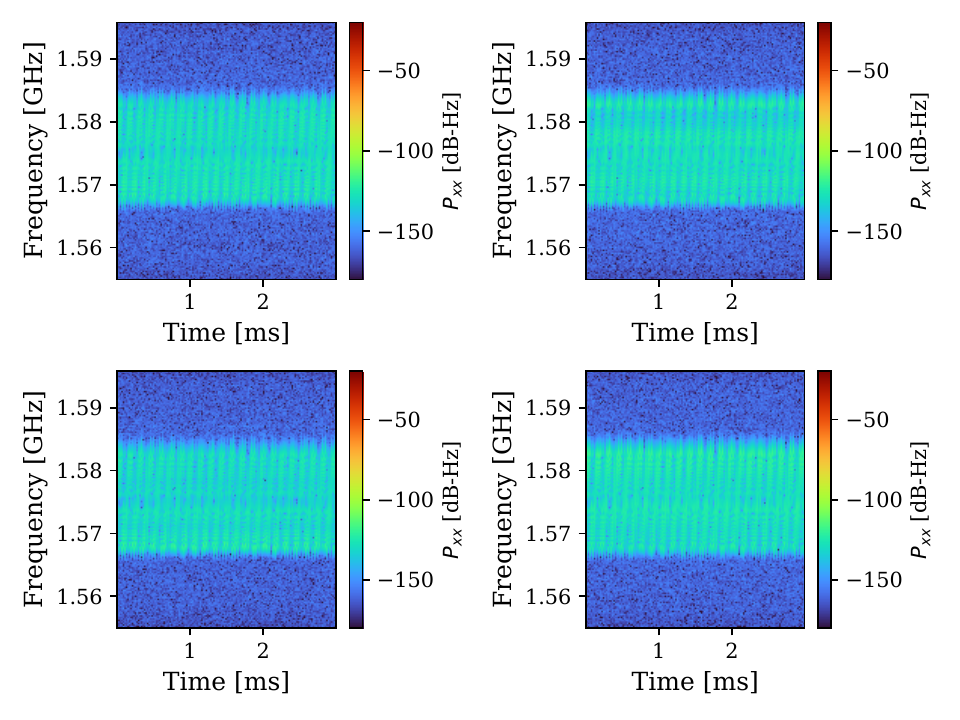}
        \caption{Spectrogram plots for the first jammer (blue) for all four antennas.}
        \label{figure_spect1}
    \end{minipage}
    \hfill
	\begin{minipage}[t]{0.492\linewidth}
        \centering
        \includegraphics[width=1.0\linewidth]{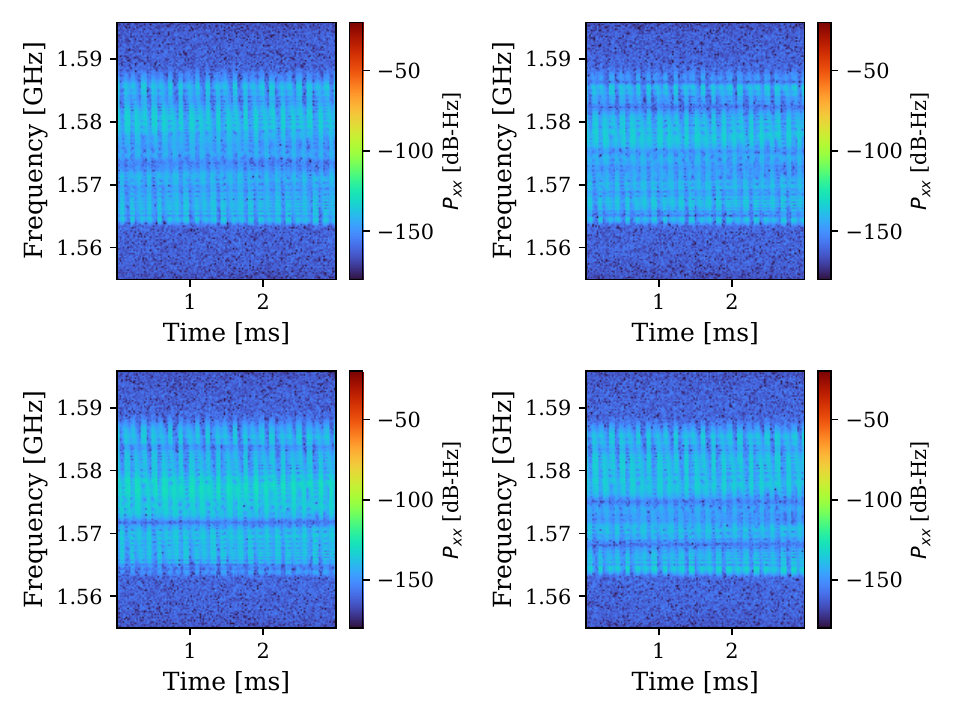}
        \caption{Spectrogram plots for the second jammer (black) for all four antennas.}
        \label{figure_spect2}
    \end{minipage}
\end{figure*}

\begin{figure*}[!t]
    \centering
	\begin{minipage}[t]{0.492\linewidth}
        \centering
        \includegraphics[width=1.0\linewidth]{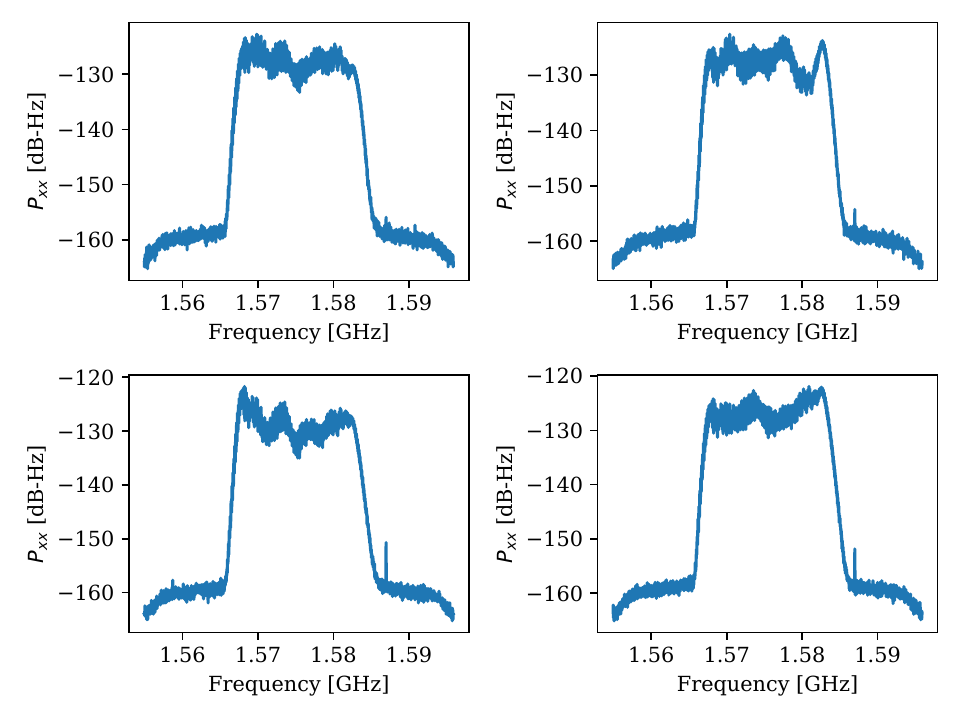}
        \caption{Welch plots for the first jammer (blue) for all four antennas.}
        \label{figure_welch1}
    \end{minipage}
    \hfill
	\begin{minipage}[t]{0.492\linewidth}
        \centering
        \includegraphics[width=1.0\linewidth]{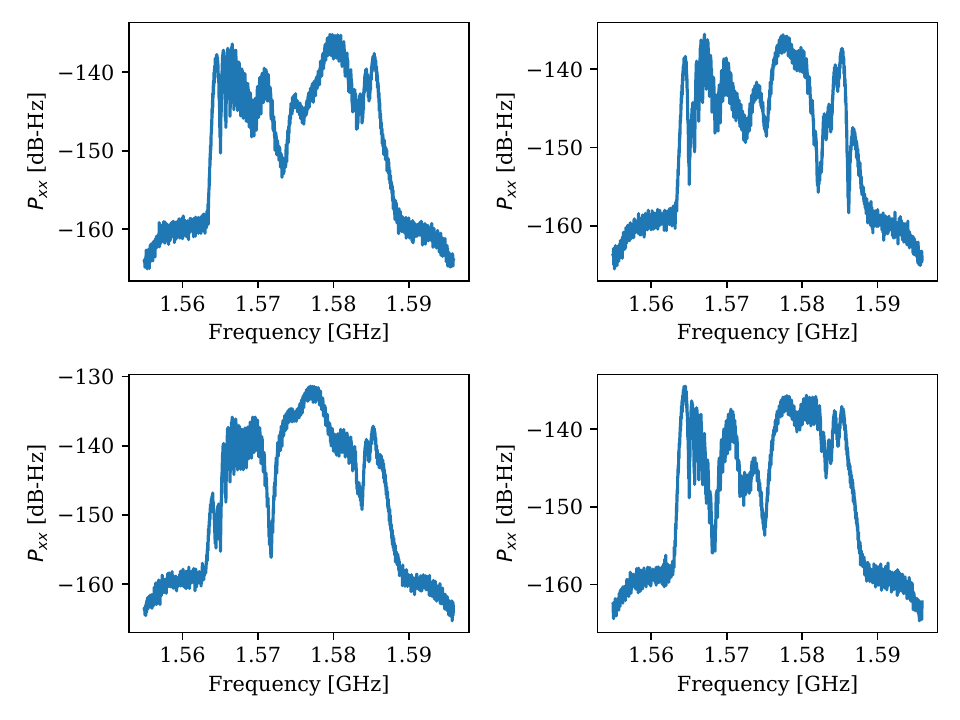}
        \caption{Welch plots for the second jammer (black) for all four antennas.}
        \label{figure_welch2}
    \end{minipage}
\end{figure*}

Subsequently, we record a series of circular trajectories consisting of $4 \times 6$ circles for each of the two jamming devices (see Figure~\ref{figure_trajectory1} and Figure~\ref{figure_trajectory2}). For each trajectory in the $4 \times 6$ grid, five concentric circles with radii ranging from $0.8\,m$ to $2\,m$ are executed. These circular trajectories are further recorded at four distinct heights -- $4.8\,m$, $4.3\,m$, $3.8\,m$, and $3.3\,m$ -- providing a comprehensive three-dimensional sampling of the measurement environment (see Figure~\ref{figure_trajectory2} and Figure~\ref{figure_trajectory4}).

The 3D positioning system provides ground truth labels, including the relative $x$-, $y$-, and $z$-coordinates between the antenna platform and the jamming device, the platform orientation with respect to the positive $x$-axis, as well as its velocity and acceleration. The synchronization between the Ettus USRP data snapshots and the positioning system measurements exhibits a time offset of less than $100\,ms$. Consequently, the ground truth labels are interpolated to align with the recorded signal data.

\subsection{Dataset}

For each jamming device, the dataset is divided into an 80/20 split for training and testing. The blue jammer dataset comprises 216,076 raw IQ samples, while the black jammer dataset contains 216,214 samples. Each sample includes four channels corresponding to the four antennas, 122,880 timesteps, and separate real and imaginary components, representing a total recording duration of 3 ms. To reduce computational load, model inputs are limited to 1,024 timesteps, equivalent to $25\,\mu s$, which retains sufficient temporal information for analysis.

Figures~\ref{figure_iq1} and \ref{figure_iq2} show example plots of the raw IQ samples and their magnitudes for both devices. The corresponding spectrograms, shown in Figure~\ref{figure_spect1} and Figure~\ref{figure_spect2}, clearly illustrate the chirp interference patterns. Differences in signal strength among the four antennas are evident, particularly in the lower-left antenna, and these variations are further highlighted in the Welch plots presented in Figure~\ref{figure_welch1} and Figure~\ref{figure_welch2}. Welch plots provide an estimate of the power spectral density (PSD) by averaging periodograms over overlapping segments, reducing variance and offering a smooth spectral representation. Together, the time-domain IQ samples, spectrograms, and Welch plots provide complementary views of the interference signals, supporting both visualization and model development for direction-finding and localization tasks.

\begin{table*}[t!]
\begin{center}
    \caption{Overview of all interference direction finding results for the first jamming device. \textbf{Bold} denotes best results (MAE) with respect to its baseline comparison.}
    \label{table_results}
    \normalsize \begin{tabular}{ p{0.5cm} | p{0.5cm} | p{0.5cm} | p{0.5cm} | p{0.5cm} }
    \multicolumn{1}{c|}{\textbf{Method}} & \multicolumn{1}{c|}{\textbf{Data Input}} & \multicolumn{1}{c}{\textbf{Distance}} & \multicolumn{1}{c}{\textbf{Azimuth}} & \multicolumn{1}{c}{\textbf{Elevation}} \\
    \multicolumn{1}{c|}{} & \multicolumn{1}{c|}{} & \multicolumn{1}{c}{\textbf{error [m]}} & \multicolumn{1}{c}{\textbf{error $\alpha [^{\circ}]$}} & \multicolumn{1}{c}{\textbf{error $\beta [^{\circ}]$}} \\ \hline
    \multicolumn{1}{l|}{HuggingFace: VovNet (30 epochs)} & \multicolumn{1}{l|}{FFT} & \multicolumn{1}{r}{1.19} & \multicolumn{1}{r}{21.37} & \multicolumn{1}{r}{4.90} \\
    \multicolumn{1}{l|}{HuggingFace: DPN68B} & \multicolumn{1}{l|}{FFT} & \multicolumn{1}{r}{0.33} & \multicolumn{1}{r}{4.39} & \multicolumn{1}{r}{2.09} \\
    \multicolumn{1}{l|}{HuggingFace: DPN48B} & \multicolumn{1}{l|}{FFT} & \multicolumn{1}{r}{0.32} & \multicolumn{1}{r}{4.37} & \multicolumn{1}{r}{2.28} \\
    \multicolumn{1}{l|}{HuggingFace: DPN68} & \multicolumn{1}{l|}{FFT} & \multicolumn{1}{r}{0.32} & \multicolumn{1}{r}{4.32} & \multicolumn{1}{r}{2.10} \\
    \multicolumn{1}{l|}{HuggingFace: EfficientNet} & \multicolumn{1}{l|}{FFT} & \multicolumn{1}{r}{\textbf{0.29}} & \multicolumn{1}{r}{\textbf{3.57}} & \multicolumn{1}{r}{\textbf{1.56}} \\ \hline
    \multicolumn{1}{l|}{tsai: TCN} & \multicolumn{1}{l|}{IQ} & \multicolumn{1}{r}{0.70} & \multicolumn{1}{r}{10.13} & \multicolumn{1}{r}{4.11} \\
    \multicolumn{1}{l|}{tsai: ResNet} & \multicolumn{1}{l|}{IQ} & \multicolumn{1}{r}{0.73} & \multicolumn{1}{r}{8.34} & \multicolumn{1}{r}{3.46} \\
    \multicolumn{1}{l|}{tsai: gMLP} & \multicolumn{1}{l|}{IQ} & \multicolumn{1}{r}{0.56} & \multicolumn{1}{r}{7.40} & \multicolumn{1}{r}{3.24} \\
    \multicolumn{1}{l|}{tsai: InceptionTime} & \multicolumn{1}{l|}{IQ} & \multicolumn{1}{r}{0.55} & \multicolumn{1}{r}{6.41} & \multicolumn{1}{r}{3.24} \\
    \multicolumn{1}{l|}{tsai: XceptionTime (50 epochs)} & \multicolumn{1}{l|}{IQ} & \multicolumn{1}{r}{\textbf{0.39}} & \multicolumn{1}{r}{\textbf{3.72}} & \multicolumn{1}{r}{\textbf{1.96}} \\
    \multicolumn{1}{l|}{tsai: XceptionTime (30 epochs)} & \multicolumn{1}{l|}{IQ} & \multicolumn{1}{r}{0.60} & \multicolumn{1}{r}{5.95} & \multicolumn{1}{r}{2.50} \\ \hline
    \multicolumn{1}{l|}{McAFF} & \multicolumn{1}{l|}{IQ} & \multicolumn{1}{r}{1.33} & \multicolumn{1}{r}{25.53} & \multicolumn{1}{r}{4.71} \\
    \multicolumn{1}{l|}{} & \multicolumn{1}{l|}{CFO} & \multicolumn{1}{r}{1.70} & \multicolumn{1}{r}{31.40} & \multicolumn{1}{r}{4.65} \\
    \multicolumn{1}{l|}{} & \multicolumn{1}{l|}{FFT} & \multicolumn{1}{r}{1.15} & \multicolumn{1}{r}{21.87} & \multicolumn{1}{r}{4.69} \\
    \multicolumn{1}{l|}{} & \multicolumn{1}{l|}{STFT} & \multicolumn{1}{r}{1.24} & \multicolumn{1}{r}{24.23} & \multicolumn{1}{r}{4.76} \\
    \multicolumn{1}{l|}{} & \multicolumn{1}{l|}{IQ+CFO+STFT} & \multicolumn{1}{r}{0.41} & \multicolumn{1}{r}{6.42} & \multicolumn{1}{r}{2.93} \\
    \multicolumn{1}{l|}{} & \multicolumn{1}{l|}{IQ+FFT+CFO+STFT} & \multicolumn{1}{r}{\textbf{0.40}} & \multicolumn{1}{r}{\textbf{6.24}} & \multicolumn{1}{r}{\textbf{2.91}} \\ \hline
    \multicolumn{1}{l|}{Fusion (VovNet + TCN)} & \multicolumn{1}{l|}{IQ+FFT} & \multicolumn{1}{r}{\textbf{0.31}} & \multicolumn{1}{r}{\textbf{3.87}} & \multicolumn{1}{r}{\textbf{2.18}} \\
    \multicolumn{1}{l|}{} & \multicolumn{1}{l|}{IQ+FFT+AoA} & \multicolumn{1}{r}{0.32} & \multicolumn{1}{r}{3.90} & \multicolumn{1}{r}{2.12} \\ \hline
    \multicolumn{1}{l|}{tsai: XceptionTime (30 epochs)} & \multicolumn{1}{l|}{IQ + relative pose (5 timesteps)} & \multicolumn{1}{r}{\textbf{0.41}} & \multicolumn{1}{r}{\textbf{3.978}} & \multicolumn{1}{r}{\textbf{1.99}} \\
    \end{tabular}
\end{center}
\end{table*}

\section{Evaluation}
\label{label_evaluation}

Hardware Setup. For all experiments, we use Nvidia Tesla V100-SXM2 GPUs with 32 GB VRAM equipped with Core Xeon CPUs and 192\,GB RAM. We use the SGD optimizer with a multi-step learning rate of $10^{-2}$, decay of $10^{-4}$, momentum of 0.9, a batch size of 64, and train for 50 epochs. We present the mean absolute error (MAE) in m and degrees ($^{\circ}$).

\textbf{Baseline Evaluation.} Table~\ref{table_results} presents a summary of all evaluation results. Initially, all Hugging Face models were benchmarked using FFT data, and the top five architectures were reported. Among these, EfficientNet achieved the lowest errors, with $\alpha = 3.57^{\circ}$ and $\beta = 1.56^{\circ}$. Subsequently, benchmarking 18 \textit{tsai} models on IQ samples identified XceptionTime as the best-performing architecture, yielding errors of $\alpha = 3.72^{\circ}$ and $\beta = 1.96^{\circ}$. The initial training was conducted for only 30 epochs, after which it was determined that a greater number of 50 training epochs was necessary for improved performance. The fusion baseline, McAFF, achieved errors of $\alpha = 6.24^{\circ}$ and $\beta = 2.91^{\circ}$ when combining all data representations (IQ, FFT, CFO, and STFT).

\begin{figure*}[!t]
    \centering
	\begin{minipage}[t]{0.325\linewidth}
        \centering
        \includegraphics[width=1.0\linewidth]{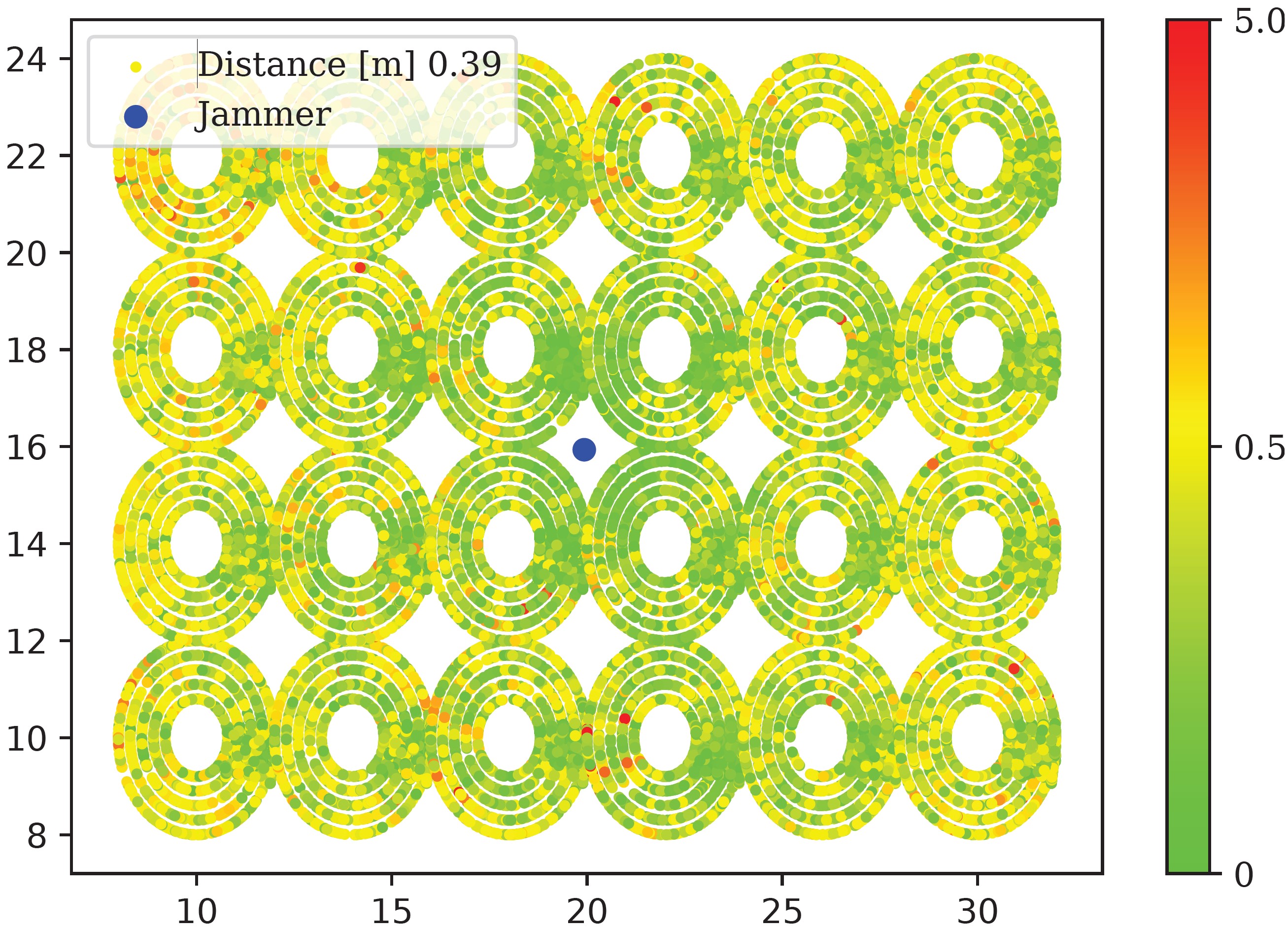}
        \caption{Distance evaluation of our fusion method (IQ+FFT+AoA).}
        \label{figure_eval_traj1}
    \end{minipage}
    \hfill
	\begin{minipage}[t]{0.325\linewidth}
        \centering
        \includegraphics[width=1.0\linewidth]{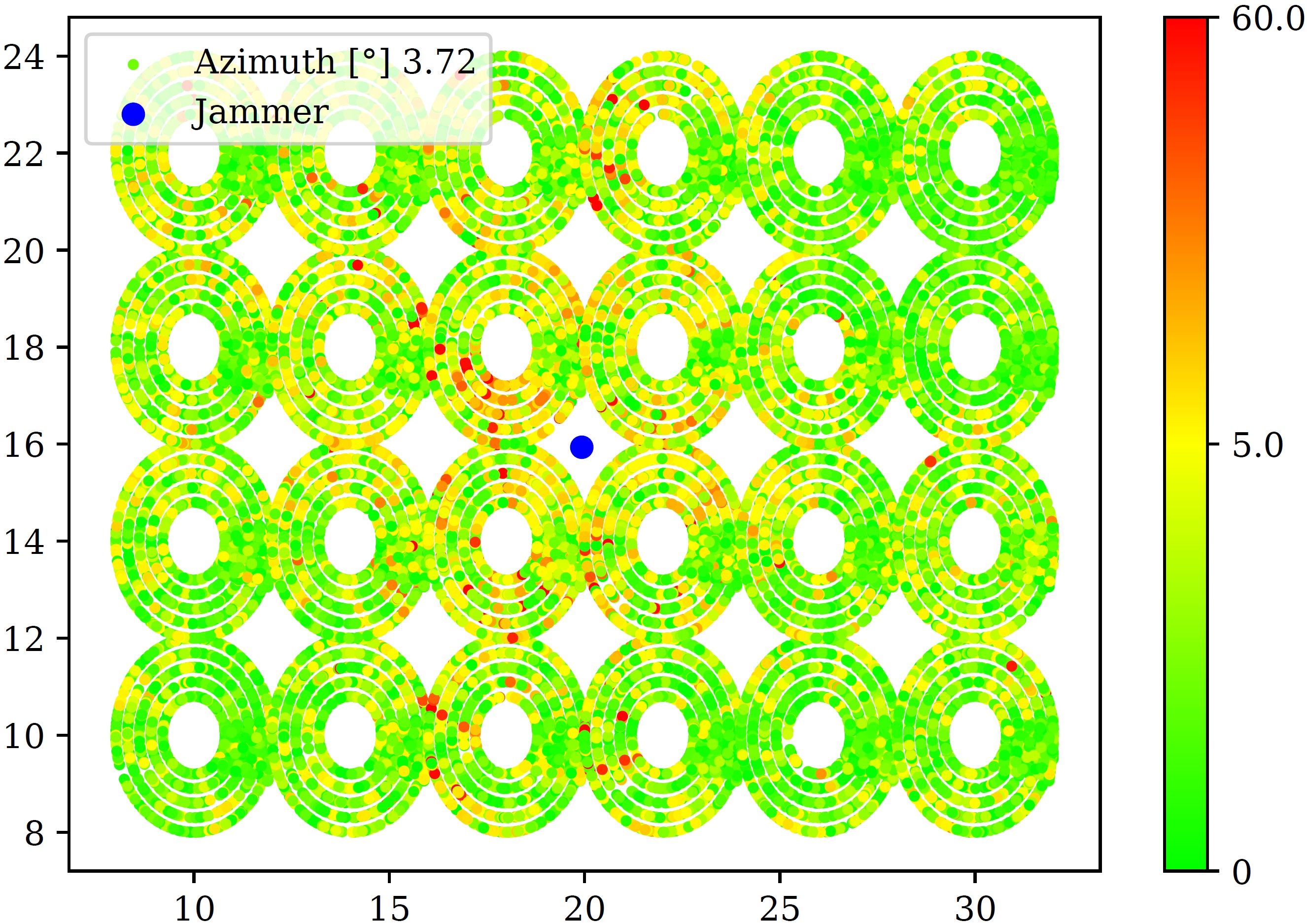}
        \caption{Azimuth evaluation of our fusion method (IQ+FFT+AoA).}
        \label{figure_eval_traj2}
    \end{minipage}
    \hfill
	\begin{minipage}[t]{0.325\linewidth}
        \centering
        \includegraphics[width=1.0\linewidth]{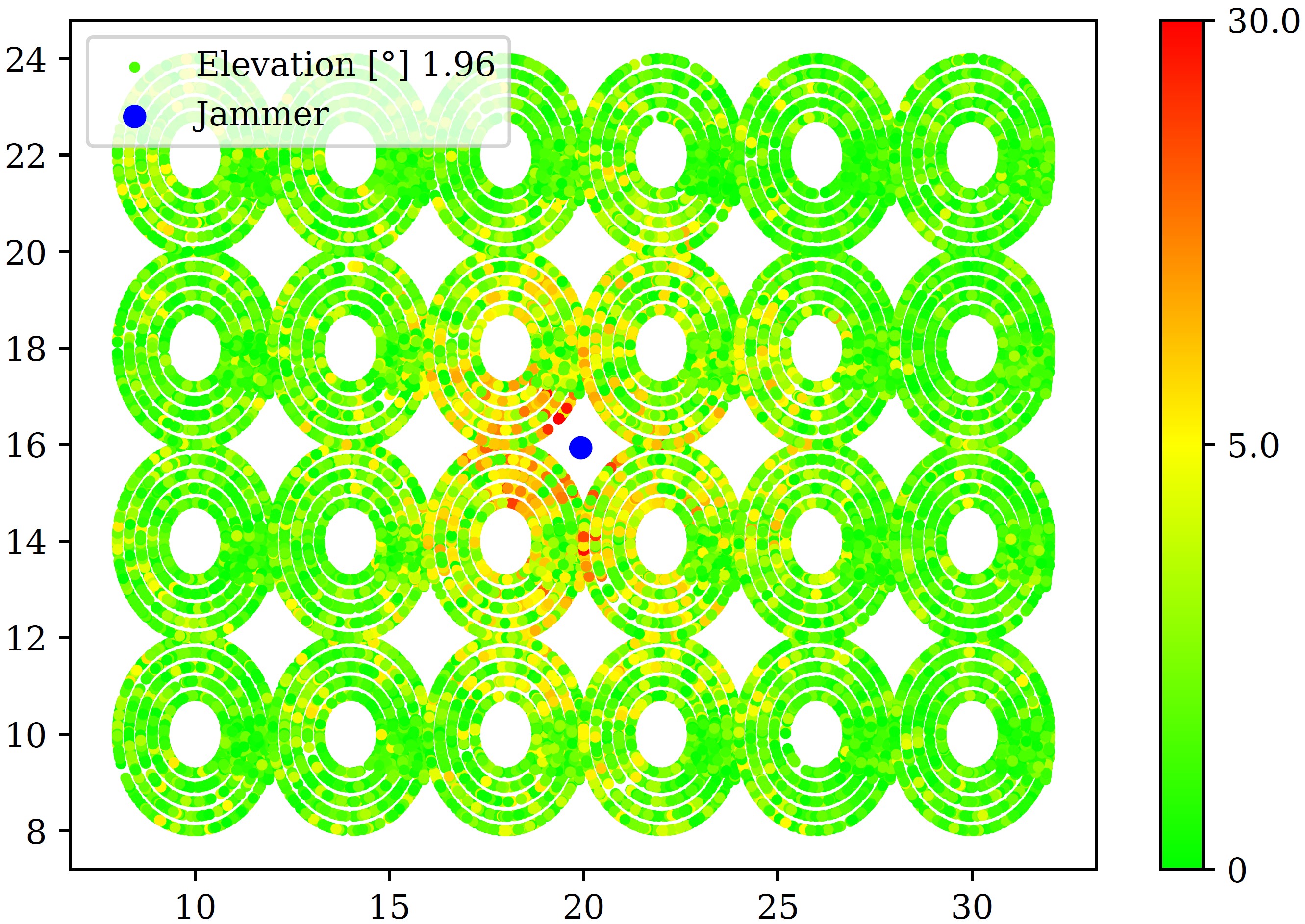}
        \caption{Elevation evaluation of our fusion method (IQ+FFT+AoA).}
        \label{figure_eval_traj3}
    \end{minipage}
\end{figure*}

\begin{figure*}[!t]
    \centering
	\begin{minipage}[t]{0.235\linewidth}
        \centering
        \includegraphics[width=1.0\linewidth]{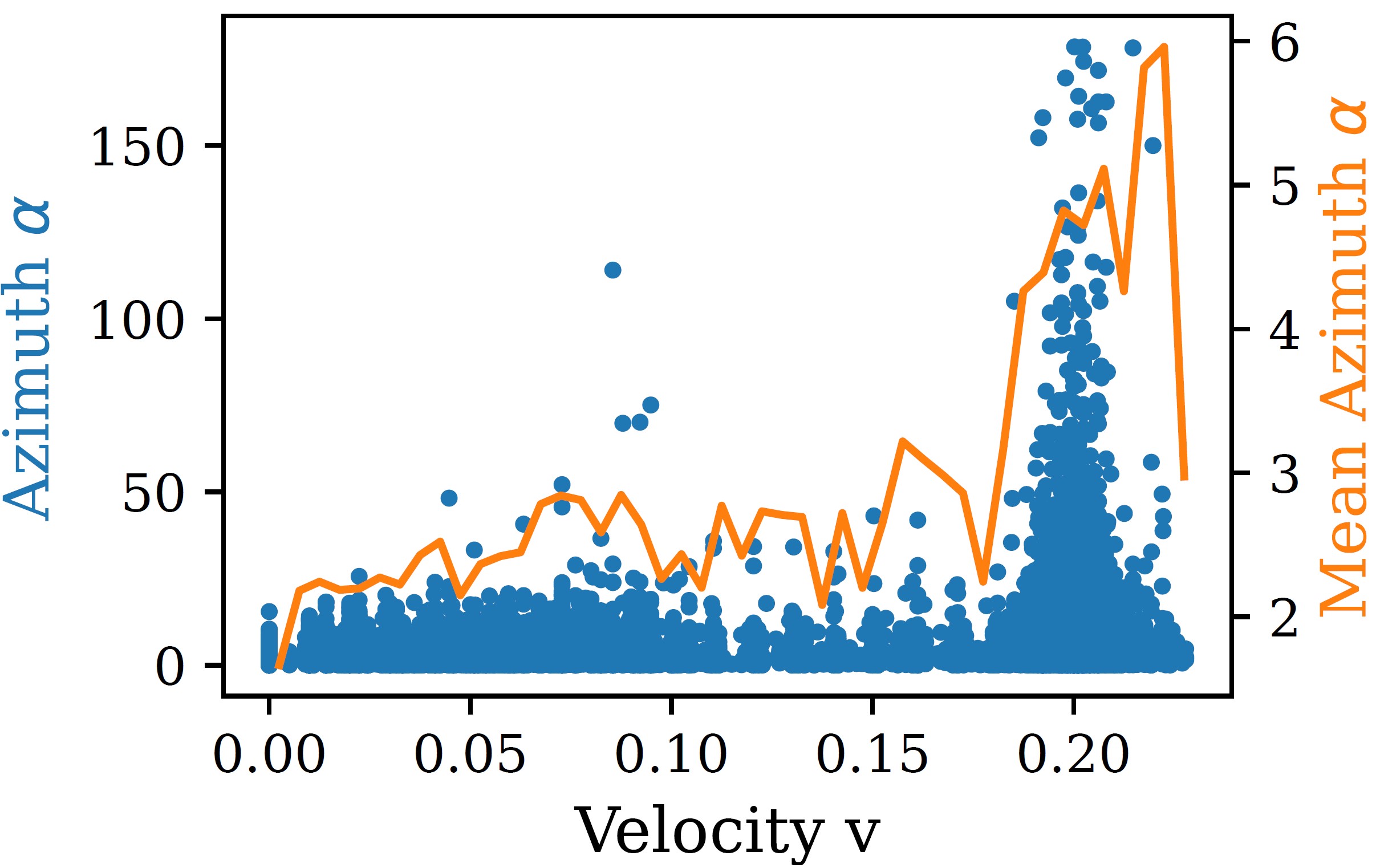}
        \caption{Correlation between azimuth error and SDR velocity.}
        \label{figure_distances1}
    \end{minipage}
    \hfill
	\begin{minipage}[t]{0.235\linewidth}
        \centering
        \includegraphics[width=1.0\linewidth]{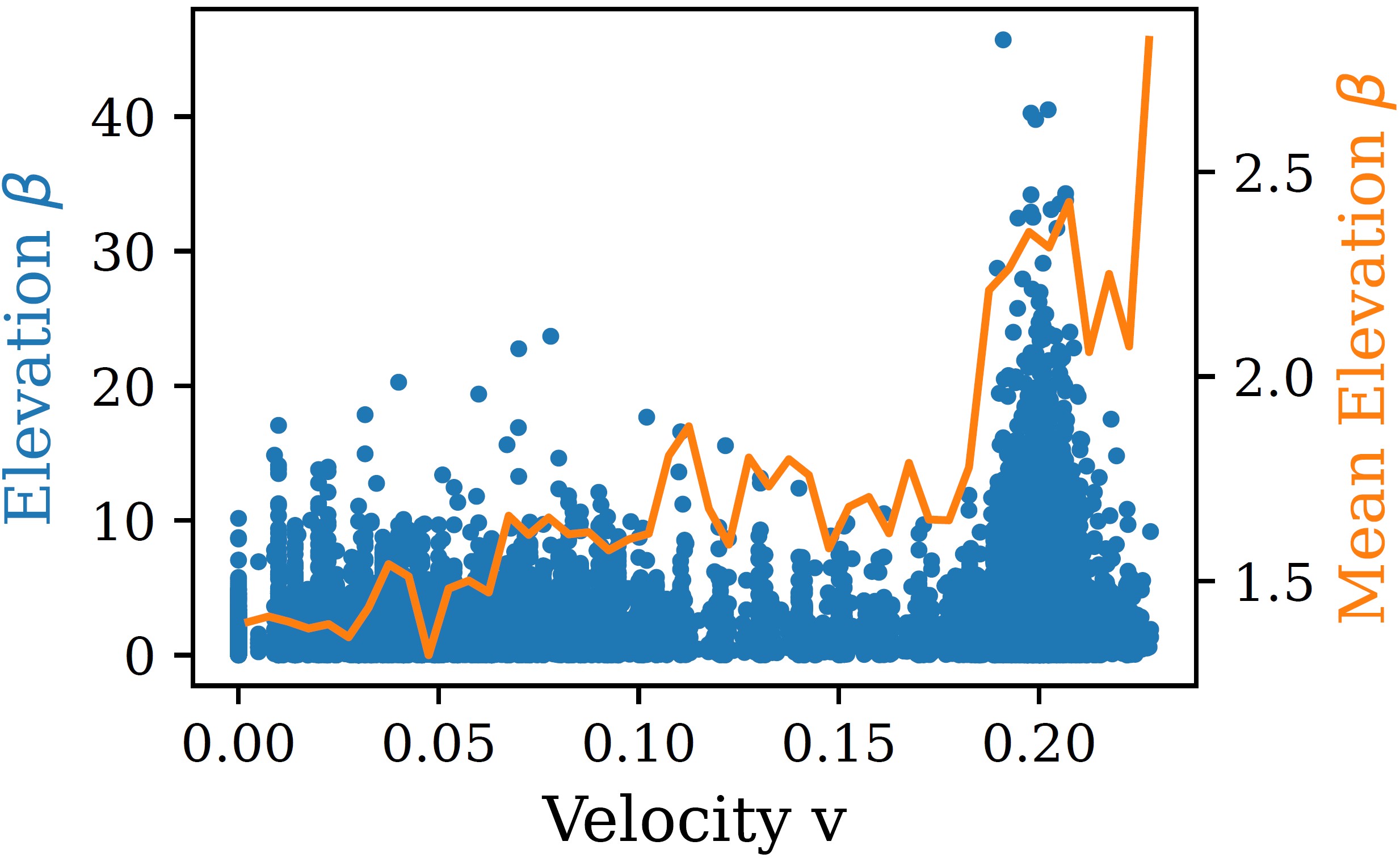}
        \caption{Correlation between elevation error and SDR velocity.}
        \label{figure_distances2}
    \end{minipage}
    \hfill
	\begin{minipage}[t]{0.235\linewidth}
        \centering
        \includegraphics[width=1.0\linewidth]{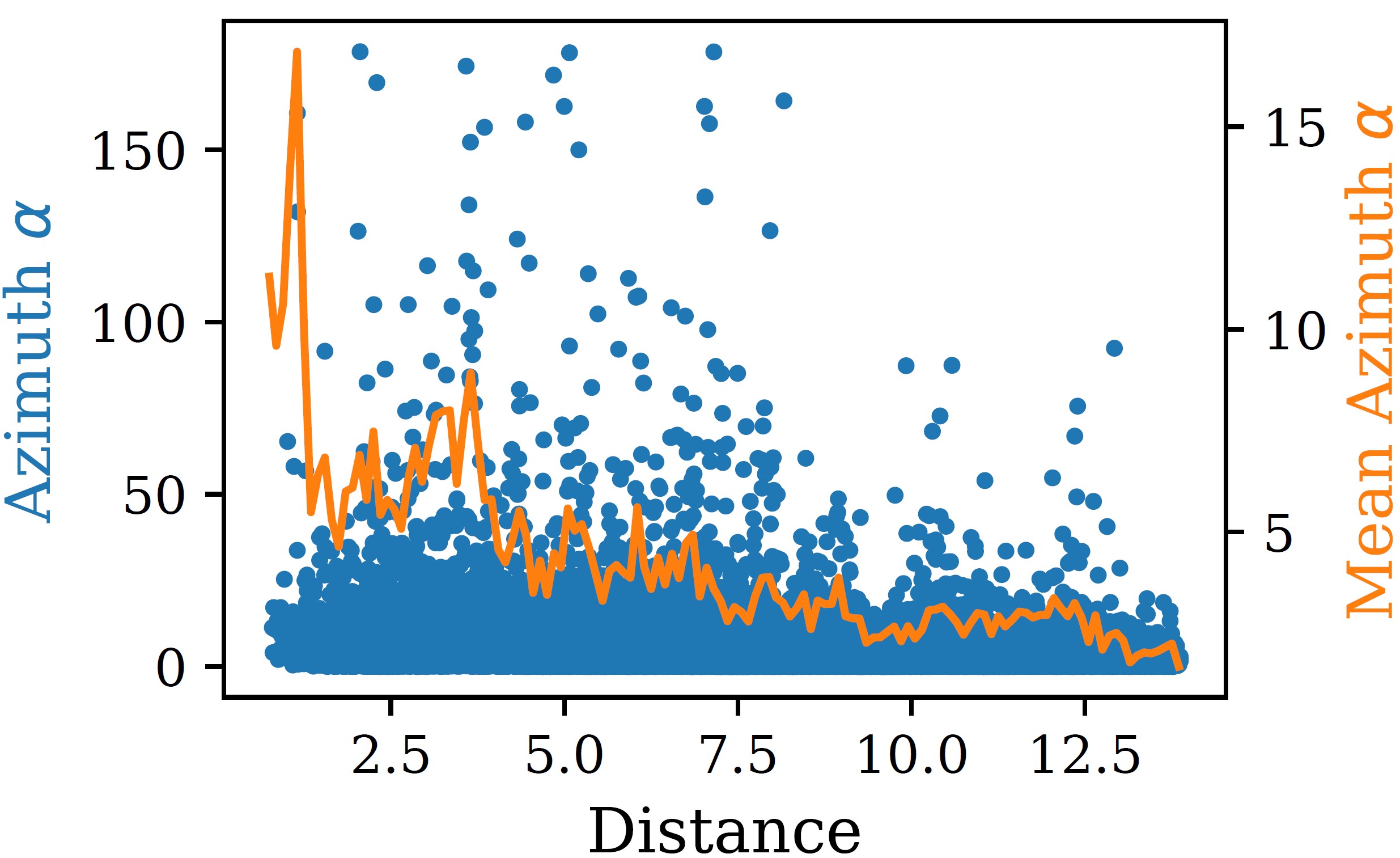}
        \caption{Correlation between azimuth error and distance to the jammer.}
        \label{figure_distances3}
    \end{minipage}
    \hfill
	\begin{minipage}[t]{0.235\linewidth}
        \centering
        \includegraphics[width=1.0\linewidth]{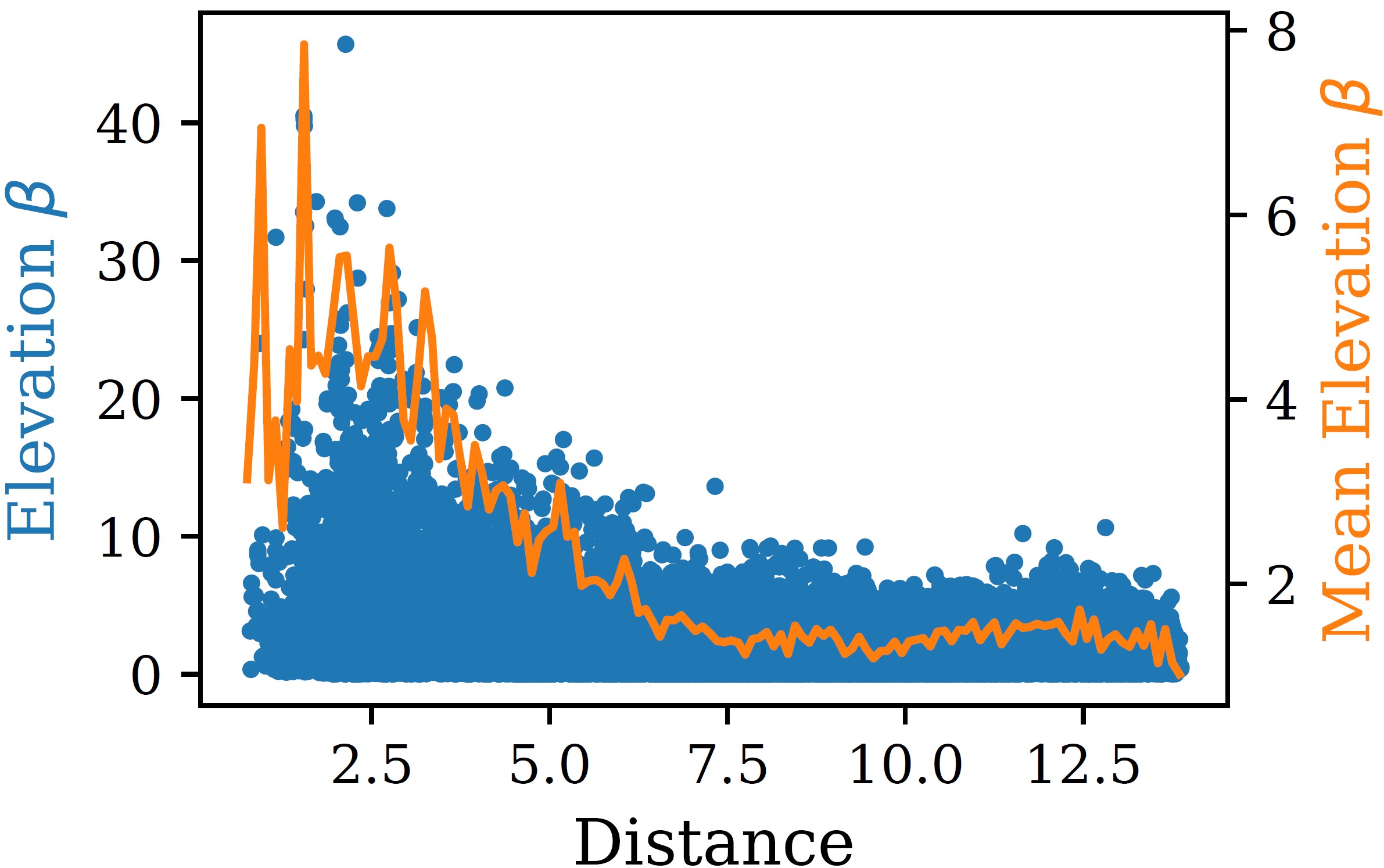}
        \caption{Correlation between elevation error and distance to the jammer.}
        \label{figure_distances4}
    \end{minipage}
\end{figure*}

\textbf{Fusion Evaluation.} This outcome motivated the fusion of FFT representations extracted with VovNet and IQ representations extracted with TCN, which achieved errors of $\alpha = 3.87^{\circ}$ and $\beta = 2.18^{\circ}$, representing a significant improvement over the individual VovNet-only and TCN-only models. Incorporating AoA statistics into the architecture led to a marginal decrease in distance and azimuth errors, while the elevation error showed a more pronounced improvement. Due to the extended training times required, the relative pose was integrated only into the XceptionTime model and trained for 30 epochs, resulting in a substantial performance enhancement. In future work, we plan to train a compressed fusion model with integrated relative poses and further investigate the optimal number of input timesteps. Figures~\ref{figure_eval_traj1} to \ref{figure_eval_traj3} present a detailed evaluation of the distance, azimuth, and elevation prediction errors for each data point obtained using the proposed fusion method. As shown in Figure~\ref{figure_eval_traj1}, the distance error increases significantly with larger separations between the antenna setup and the interference source. Conversely, for shorter distances, the azimuth error tends to increase, primarily because the positioning system is located at a higher elevation than the interference source (see Figure~\ref{figure_eval_traj2}). Additionally, due to the antenna pattern characteristics, the interference source lies outside the main angle of incidence, further contributing to this error. Similarly, and more distinctly, Figure~\ref{figure_eval_traj3} illustrates that the elevation error decreases as the distance between the antennas and the interference source increases.

\begin{figure}[!t]
    \centering
    \includegraphics[trim=15 15 15 15, clip, width=0.8\linewidth]{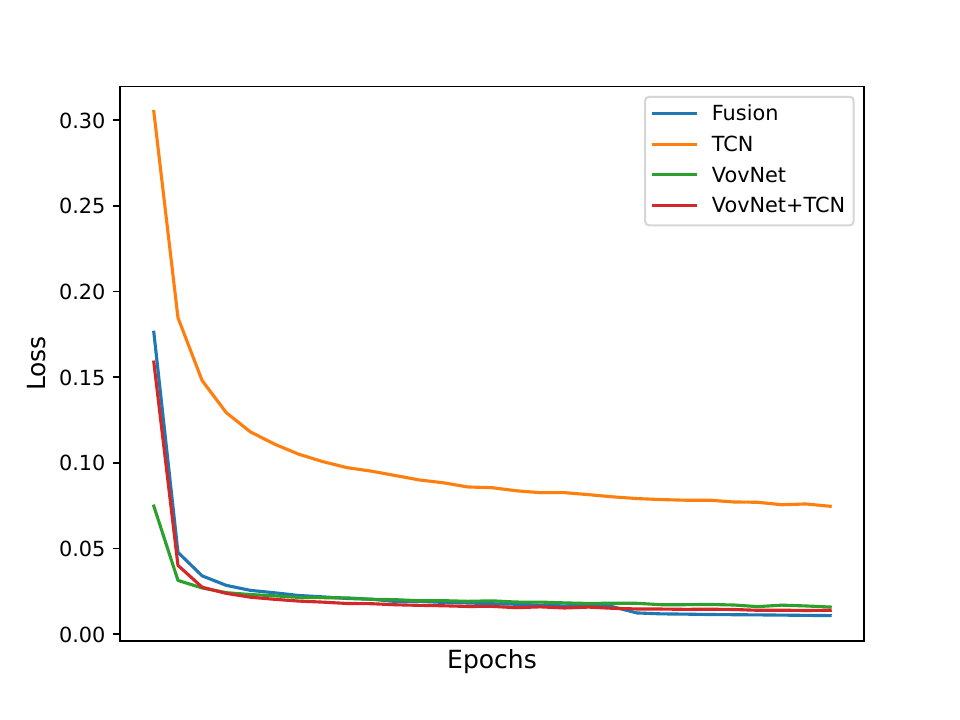}
    \caption{Comparison of the loss for the TCN, VoVNet, VovNet+TCN, and Fusion models.}
    \label{figure_loss_evaluation}
\end{figure}

\textbf{Loss Evaluation.} Figure~\ref{figure_loss_evaluation} presents a comparison of the training loss for the TCN-only, VovNet-only, VovNet+TCN combination, and the proposed fusion model integrating IQ, FFT, and AoA features. Among these, the TCN model exhibits the highest loss, whereas the VovNet model demonstrates faster convergence. Combining both architectures further reduces the training loss and consequently decreases the prediction errors. The proposed fusion approach, which additionally incorporates AoA statistics, yields a marginal yet consistent improvement in loss convergence.

\begin{table*}[t!]
\begin{center}
    \caption{Cross-validation of the two jamming devices for the XceptionTime model.}
    \label{table_generalization_jammer}
    \normalsize \begin{tabular}{ p{0.5cm} | p{0.5cm} | p{0.5cm} | p{0.5cm} | p{0.5cm} | p{0.5cm} | p{0.5cm} }
    & \multicolumn{3}{c|}{\textbf{Jammer 1}} & \multicolumn{3}{c}{\textbf{Jammer 2}} \\
    & \multicolumn{1}{c}{\textbf{Distance}} & \multicolumn{1}{c}{\textbf{Azimuth}} & \multicolumn{1}{c|}{\textbf{Elevation}} & \multicolumn{1}{c}{\textbf{Distance}} & \multicolumn{1}{c}{\textbf{Azimuth}} & \multicolumn{1}{c}{\textbf{Elevation}} \\
    & \multicolumn{1}{c}{\textbf{error [m]}} & \multicolumn{1}{c}{\textbf{error $\alpha [^{\circ}]$}} & \multicolumn{1}{c|}{\textbf{error $\beta [^{\circ}]$}} & \multicolumn{1}{c}{\textbf{error [m]}} & \multicolumn{1}{c}{\textbf{error $\alpha [^{\circ}]$}} & \multicolumn{1}{c}{\textbf{error $\beta [^{\circ}]$}} \\ \hline
    \multicolumn{1}{c|}{\textbf{Jammer 1}} & \multicolumn{1}{r}{0.39} & \multicolumn{1}{r}{3.72} & \multicolumn{1}{r|}{1.96} & \multicolumn{1}{r}{3.89} & \multicolumn{1}{r}{94.07} & \multicolumn{1}{r}{5.60} \\
    \end{tabular}
\end{center}
\end{table*}

\textbf{Generalization.} We next evaluate the generalization capability of the model. In Table~\ref{table_results}, only the results for the first jamming device were presented, as the outcomes for the second jammer were initially observed to be comparable. However, when performing cross-validation between both jamming devices (see Table~\ref{table_generalization_jammer}), the performance notably degrades, with the azimuth error increasing from $\alpha = 3.72^{\circ}$ to $\alpha = 94.07^{\circ}$ and the elevation error increasing from $\beta = 1.96^{\circ}$ to $\beta = 5.60^{\circ}$. This substantial rise indicates a reduced generalization capability when the model is applied to data from an unseen jammer. To further investigate the underlying causes, Figures~\ref{figure_distances1} and \ref{figure_distances2} analyze the correlation between azimuth and elevation errors and the SDR velocity. The mean trend, depicted in orange, shows that higher velocities (around 0.20 m/s) lead to a marked increase in azimuth error, reaching approximately $\alpha = 6^{\circ}$ on average. In contrast, the elevation error exhibits a less pronounced dependency on velocity, with an average increase of up to $\beta = 3^{\circ}$. Figures~\ref{figure_distances3} and \ref{figure_distances4} illustrate the relationship between prediction errors and the distance between the antenna system and the interference source. Consistent with previous observations (see Figures~\ref{figure_eval_traj2} and \ref{figure_eval_traj3}), both azimuth and elevation errors increase significantly at shorter distances (less than $7.5\,m$). This behavior suggests that proximity to the interference source amplifies angular estimation inaccuracies, likely due to limitations in the antenna’s spatial resolution and the geometry of the measurement setup.
\section{Conclusion}
\label{label_conclusion}

This work introduced a novel multimodal fusion architecture for GNSS interference localization that integrates time-series IQ data, FFT-based spectrogram features, conventional AoA statistics, and IMU-derived relative pose information. The proposed system, built on a large-scale dataset recorded using a high-precision SDR platform within a controlled industrial environment, demonstrates superior performance over state-of-the-art baselines such as McAFF. In particular, the combined VovNet+TCN fusion model achieved a minimum azimuth error of $\alpha = 3.87^{\circ}$ and elevation error of $\beta = 2.18^{\circ}$, marking a substantial improvement over single-modality approaches. The inclusion of AoA features and motion-based cues further enhanced convergence stability and overall accuracy. While cross-validation across different jamming devices revealed a reduction in generalization performance, the presented analysis identified key influencing factors such as platform velocity and antenna–source distance, providing valuable insights for model robustness. Overall, the proposed approach establishes an effective framework for robust and physically consistent jammer localization and forms the basis for future research on compressed fusion architectures and extended temporal modeling.

\section*{Acknowledgments}
This work has been carried out within the PaiL project, funding code 50NP2506, sponsored by the German Federal Ministry for Transport (BMV) and supported by the German Aerospace Center (DLR), the Bundesnetzagentur (BNetzA), and the Federal Agency for Cartography and Geodesy (BKG).

\bibliography{DGON_ISA2025}
\bibliographystyle{IEEEtran}

\end{document}